\documentclass[a4paper,11pt]{article}
\usepackage{jheppub} 

\usepackage[T1]{fontenc} 

\usepackage{amsmath}
\usepackage{amssymb}
\usepackage{graphicx}
\usepackage{color}
\usepackage{epsfig}
\usepackage{float}
\usepackage{soul}
\usepackage{bm}
\usepackage[section]{placeins}
\usepackage{times}
\usepackage{hyperref}
\hypersetup{
  colorlinks=true,
  citecolor=blue,
  linkcolor=blue,
  urlcolor=blue}

\usepackage{epsfig}
\usepackage{dcolumn}
\usepackage{morefloats}
\usepackage[dvipsnames]{xcolor}

\usepackage{caption}
\usepackage{graphicx}
\usepackage{mathtools}
\usepackage{color}
\usepackage{braket}
\usepackage{hyperref}
\usepackage{url}
\usepackage{multirow}
\usepackage{relsize}
\usepackage{makecell}
\usepackage{float}
\usepackage{bm}

\def\bea{\begin{eqnarray}}
\def\eea{\end{eqnarray}}
 \def\be{\begin{equation}}
\def\ee{\end{equation}}

\newcommand{\dms}{\mbox{$\Delta m^2_{\odot}$}}

\def\gsim{\ \rlap{\raise 2pt\hbox{$>$}}{\lower 2pt \hbox{$\sim$}}\ }
\def\lsim{\ \rlap{\raise 2pt\hbox{$<$}}{\lower 2pt \hbox{$\sim$}}\ }
\def\dslash{\kern-4pt \not{\hbox{\kern-2pt $\partial$}}}
\def\pslash{\not{\hbox{\kern-2pt p}}}

\def\dam{\mathrm{(\Delta m_{31}^2)^m}}

\def\da{\mathrm{\Delta m_{31}^2}}

\def\dcp{{\delta_{\mathrm{CP}}}}
\def\stsmallm{\mathrm{\sin^2 2 \theta_{13}^m}}
\newcommand{\nova}{NO$\nu$A}

\def \nn {\nonumber}
\def \nova {NO$\nu$A }
\def \ess {ESS$\nu$SB }
\def \t2k {T2K }
\def \dcp {$\delta_{CP}~$}
\def \th23 {$\theta_{23}~$}
\def \dms31 {$\Delta m^2_{31}~$}

\DeclareGraphicsExtensions{.eps,.ps}

\title{\boldmath Enhancing the hierarchy and octant sensitivity of 
ESS$\nu$SB in conjunction with T2K, \nova and  ICAL@INO}

\author[a,b]{Kaustav Chakraborty,}
\author[a]{Srubabati Goswami,}
\author[a,c]{Chandan Gupta,}
\author[d]{Tarak Thakore}

\affiliation[a]{Theoretical Physics Division, Physical Research Laboratory,\\Ahmedabad - 380009, India}
\affiliation[b]{Discipline of Physics, Indian Institute of Technology,\\Gandhinagar - 382355, India}
\affiliation[c]{Homi Bhabha National Institute,\\Training School Complex,
  Anushakti Nagar, Mumbai 400085, India}
\affiliation[d]{Institut de Fisica Corpuscular (CSIC-Universitat de Valencia),\\Parc Cientific de la UV C/ Catedratico Jose Beltran, 2, E-46980 Paterna (Valencia), Spain}

\emailAdd{kaustav@prl.res.in}
\emailAdd{sruba@prl.res.in}
\emailAdd{ph10c009@gmail.com}
\emailAdd{tarak.thakore@ific.uv.es}

\abstract{
The  main aim of the ESS$\nu$SB proposal 
is the discovery of the leptonic CP phase \dcp  with a high significance 
($5\sigma$ for 50\% values of \dcp) 
by utilizing the physics at the second oscillation maxima of the $P_{\mu e}$
channel. It can achieve $3\sigma$ sensitivity to hierarchy for all values of 
\dcp. 
In this work, we  concentrate on the hierarchy and octant 
sensitivity of the \ess experiment. 
We 
show that combining the ESS$\nu$SB experiment 
with the atmospheric neutrino data from the proposed India-based Neutrino Observatory(INO)
experiment can result in an increased sensitivity to mass hierarchy.  
In addition, we also combine the results from the ongoing experiments 
T2K and \nova assuming their full runtime and present the combined 
sensitivity of \ess+ ICAL@INO + T2K + NO$\nu$A.
We show that while by itself ESS$\nu$SB can have up to 
$3\sigma$  hierarchy sensitivity, the combination of all
the experiments can give up to $5\sigma$ sensitivity depending on the 
true hierarchy-octant combination. 
The octant sensitivity of 
ESS$\nu$SB is low by itself. However the combined sensitivity of all the 
above experiments can give up to $3\sigma$ sensitivity depending on 
the choice of true hierarchy and octant. We discuss the various degeneracies
and the synergies that lead to the enhanced sensitivity when combining 
different experimental data. 
}

\begin{document} 

\maketitle
\flushbottom

\section{Introduction}
Several commendable ongoing and past experiments have not only 
established neutrino oscillation phenomenon, the  
oscillation parameters have also been quantified with considerable precision.  
In the three flavor framework, these are the mixing angles
$\theta_{12}$, $\theta_{13}$, $\theta_{23}$ and the mass squared differences
$\Delta m^2_{21}$, and $|\Delta m^2_{31}|$, where $\Delta m^2_{ij} = m_i^2 - m_j^2$, $i,j = 1,2,3 ~ \& ~ i >j$.
The unresolved parameters at this stage are the -- neutrino mass hierarchy or 
the sign of $\Delta m^2_{31}$, octant of $\theta_{23}$ and the CP phase $\delta_{CP}$.
The mass hierarchy or mass ordering depends on the  position of  
the mass eigenstate state $m_3$ with respect to the 
other two. If $m_3 > m_2 > m_1$, then it is called Normal Hierarchy (NH) 
whereas if $m_3 < m_1 < m_2$ it is called Inverted Hierarchy (IH). 
As for the octant,  
$\theta_{23}<45^\circ$ implies Lower 
Octant (LO) and  $\theta_{23}>45^\circ$ corresponds to 
the Higher Octant (HO). 
CP phase $\delta_{CP}$ is the parameter which governs the CP 
violation in the neutrino sector: $\delta_{CP} = 0^\circ,\pm 180^\circ$ implies CP
conservation whereas $\delta_{CP} = \pm 90^\circ$ 
corresponds to maximum CP violation. 
Global analysis of data from all the neutrino oscillation experiments
indicates \dcp $\sim  -90^\circ$, reports a  hint for 
NH and a preference for HO 
\cite{Capozzi:2017ipn,Esteban:2016qun,deSalas:2017kay}. 
One of the  main difficulties in determining these parameters 
 are the presence of 
degeneracies due to the unknown value of the phase $\delta_{CP}$.  
Several future experiments are proposed or planned to address the above 
degeneracies and unambiguous determination of the parameters -- hierarchy, octant and $\delta_{CP}$. 
This includes the beam based experiments 
T2HK \cite{Abe:2014oxa} $/$ T2HKK \cite{Abe:2016ero},
 DUNE \cite{Acciarri:2015uup} and \ess \cite{Baussan:2012cw,Baussan:2013zcy}.
Many studies have been performed in the literature to explore the physics 
potential of these facilities \cite{Agarwalla:2013hma,Ghosh:2014rna,
Bora:2014zwa,Barger:2013rha,Deepthi:2014iya,Nath:2015kjg,
C.:2014ika,Coloma:2012ji,Ballett:2016daj, Agarwalla:2017nld,Ghosh:2015tan, 
Ghosh:2017ged,Raut:2017dbh,Das:2017fcz}. A recent comparative study of these facilities 
have been accomplished in \cite{Chakraborty:2017ccm}.  
Among these, the \ess proposal plans to use the
European Spallation Source (ESS), 
which is under construction in Sweden. The \ess experiment 
will use this facility for generating a very intense neutrino 
super-beam. The main aim of the \ess experiment is to measure the CP violation
in the neutrino sector. This is expected to be achieved by using the 
second oscillation maxima of the $P_{\mu e}$ probability. 
However, for such a set-up the hierarchy and octant 
sensitivity gets compromised as compared
to the first oscillation maximum. 
Optimization of the \ess proposal for discovery of 
\dcp has been done in \cite{Baussan:2013zcy}, which recommends the neutrino baseline in the 300-550 km range and peak energy of 0.24 GeV.  
It was shown that $5\sigma$ sensitivity can be 
achieved for discovery of CP violation for 50\% of the \dcp values for 
2 years  of $\nu$ and 8 years  of $\bar{\nu}$ run. This configuration can also 
reach $3\sigma$ hierarchy sensitivity for majority of the \dcp values. 
In \cite{Agarwalla:2014tpa} the octant sensitivity of \ess has been studied at
both first and second oscillation maxima and they advocated 200 km baseline with 7$\nu$+3$\bar{\nu}$ years as the optimal configuration for octant and CP sensitivity.

In this study   
our aim is to explore whether the hierarchy and octant 
sensitivity  of  \ess 
can be improved by combining with
the proposed atmospheric neutrino experiment ICAL@INO\cite{Kumar:2017sdq}. 
It has been shown earlier that since the hierarchy sensitivity of ICAL@INO 
is independent of $\delta_{CP}$, combination of ICAL@INO with T2K and \nova can help 
to raise the hierarchy sensitivity for unfavorable values of \dcp for the 
latter experiments\cite{Chatterjee:2013qus}. 
We perform a quantitative analysis of this 
effect for the \ess experiment. In addition we also explore how the 
information from the ongoing experiments T2K and \nova can further enhance 
the hierarchy and octant sensitivity of the \ess experiment as well as
\ess+ ICAL@INO combination. 
We discuss in detail the various degeneracies 
and expound the synergistic effects of combining data from  the various 
experiments. 
The \ess set-up that we consider corresponds to that discussed in 
\cite{Baussan:2013zcy} with a baseline of 540 km -- between Lund 
and Garpenberg mine.

This paper is structured as follows. 
In section \ref{sec:Probability analysis}, we have described the 
appearance probabilities of the long-baseline accelerator experiments (LBL) and
the associated degeneracies. We also discuss briefly the  behavior of the 
probability  
for baselines  and energies for which resonance matter effect occurs for 
atmospheric neutrinos passing through the earth. 
Section \ref{sec:Experimental-details}, summarizes the 
various 
experiments that are used in this analysis and in section \ref{sec:Simulation details} the details of the simulation procedure is described.  
The section \ref{sec:Results and Discussions} contains the results for the 
mass hierarchy and octant sensitivity that can be achieved from  \ess, ICAL@INO ,
\t2k + \nova and their various combinations. 
Conclusions are presented in section \ref{sec:Conclusions}.

\section{Probability analysis}\label{sec:Probability analysis}

For the accelerator based experiments T2K, NO$\nu$A, \ess the 
relevant channel for mass hierarchy and octant sensitivity is  
the appearance channel  governed by the probability $P_{\mu e}$. 
In presence of matter of constant density, this can be expanded in terms of 
the small  parameters $\alpha~(=~\frac{\Delta m^2_{21}}{\Delta m^2_{31}})$ and $\sin{\theta_{13}}$ up to second order as \cite{Akhmedov:2004ny} 
\footnote{For a recent discussion on probability expressions 
under various approximations and more accurate expressions
see \cite{Parke:2019vbs}}:
 \begin{eqnarray}\label{eq:pmue}
  P(\nu_\mu \rightarrow \nu_e) &=& \sin^2{2\theta_{13}} \sin^2{\theta_{23}} \frac{\sin^2{\Delta} (1 - \hat{A})}{(1 - \hat{A})^2}  \nn \\
                               && + \alpha \cos{\theta_{13}} \sin{2\theta_{12}} \sin{2\theta_{13}} \sin{2\theta_{23}} \cos{(\Delta + \delta_{CP})} \frac{  \sin{\Delta\hat{A}}}{\hat{A}}     \frac{ \sin{\Delta(1-\hat{A})}}{(1-\hat{A})}  \nn \\
                               && + \alpha^2 \sin^2{2\theta_{12}} \cos^2{\theta_{13}} \cos^2\theta_{23} \frac{\sin^2{\Delta\hat{A}}}{\hat{A}^2}
 \end{eqnarray}
where, $E$ = energy of the neutrino, $ A = 2 \sqrt{2} G_F N_e E $, $G_F$ is the Fermi constant, $N_e$ is the number density of electrons in matter, $L$ is the baseline, $\Delta = \frac{\Delta m^2_{31}L}{4E}$, $\alpha = \frac{\Delta m^2_{21}}{\Delta m^2_{31}}$, $\hat{A} = \frac{A}{\Delta m^2_{31}}$, ${\Delta m^2_{ij}} = m^2_i - m^2_j$.  

\begin{figure}[h!]
\begin{center}
        \begin{tabular}{clr}
	\hspace{-3cm}
	\includegraphics[width=0.55\textwidth]{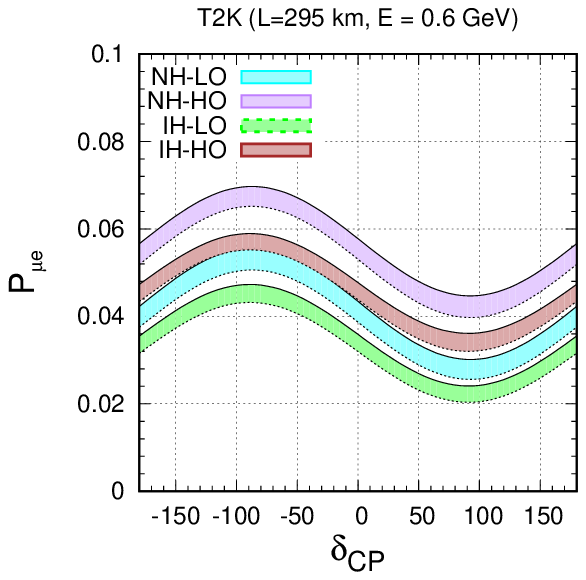}
	\hspace{-3cm}
        \includegraphics[width=0.55\textwidth]{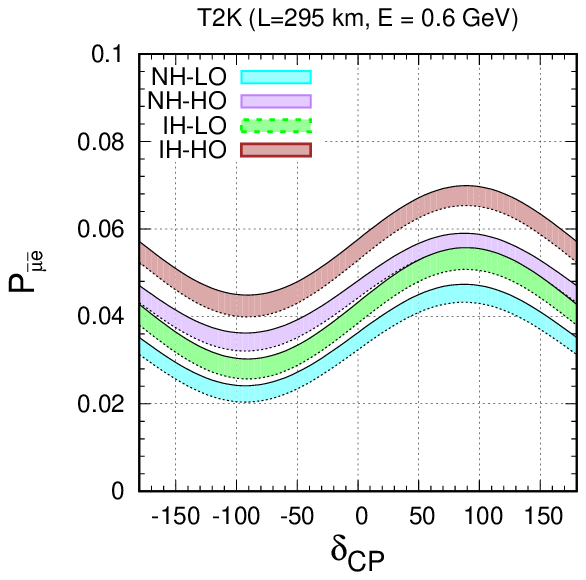}\\	
	\hspace{-3cm}
        \includegraphics[width=0.55\textwidth]{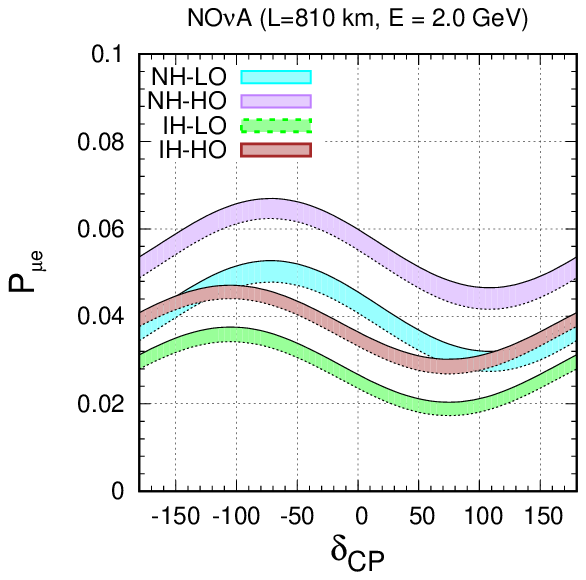}
        \hspace{-3cm}
        \includegraphics[width=0.55\textwidth]{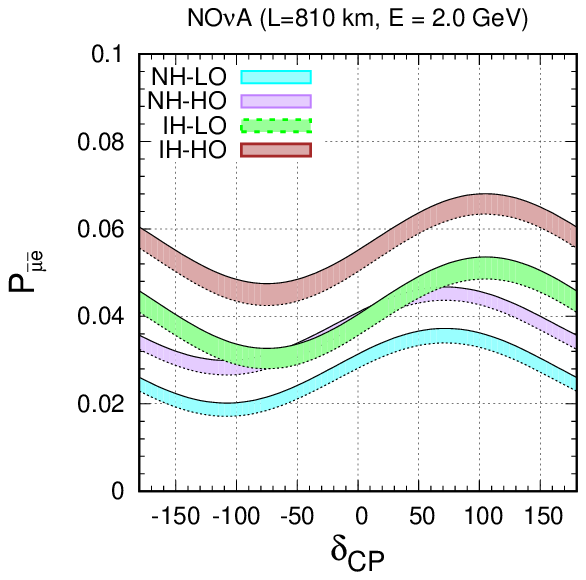}
	
\end{tabular}

\end{center}
\caption{Probability vs \dcp for \t2k \& NO$\nu$A. The bands are obtained 
by varying $\theta_{23}$ in lower octant and upper octant. See text for details.  }
\label{fig:probPlots}
\end{figure}

\begin{figure}[h!]
\begin{center}
        \begin{tabular}{cc}
        \hspace{-3cm}
        \includegraphics[width=0.55\textwidth]{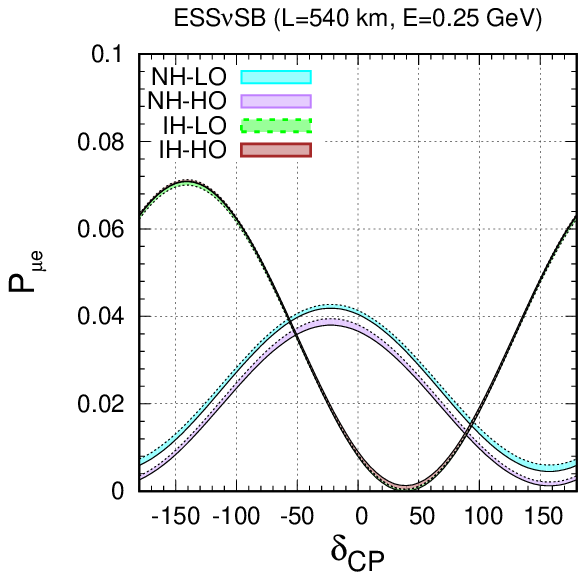}
	\hspace{-3cm}
	\includegraphics[width=0.55\textwidth]{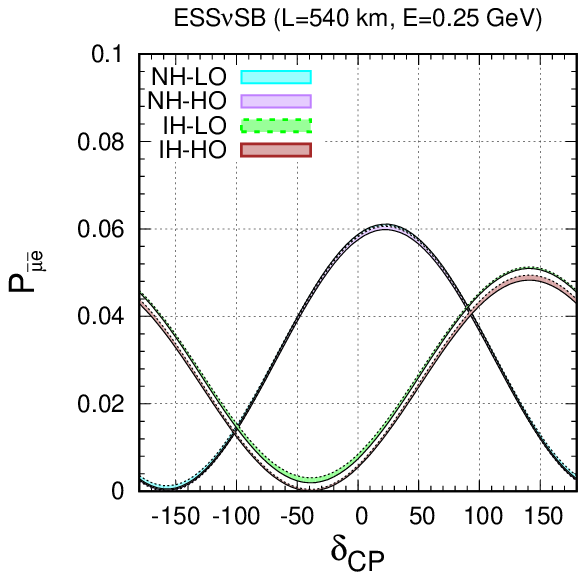}\\
         \hspace{-3cm}
         \includegraphics[width=0.55\textwidth]{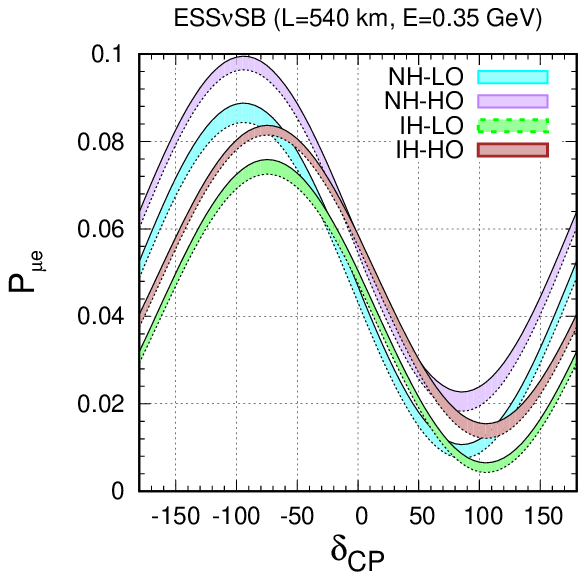}
	\hspace{-3cm}
	\includegraphics[width=0.55\textwidth]{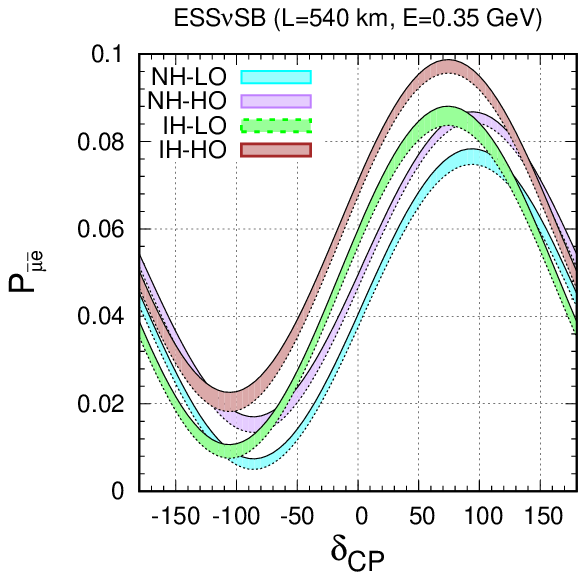}\\
	\hspace{-3cm}
         \includegraphics[width=0.55\textwidth]{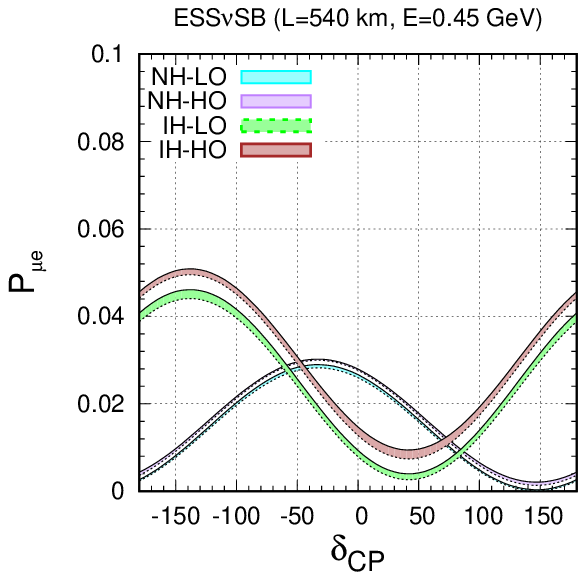}
	\hspace{-3cm}
	\includegraphics[width=0.55\textwidth]{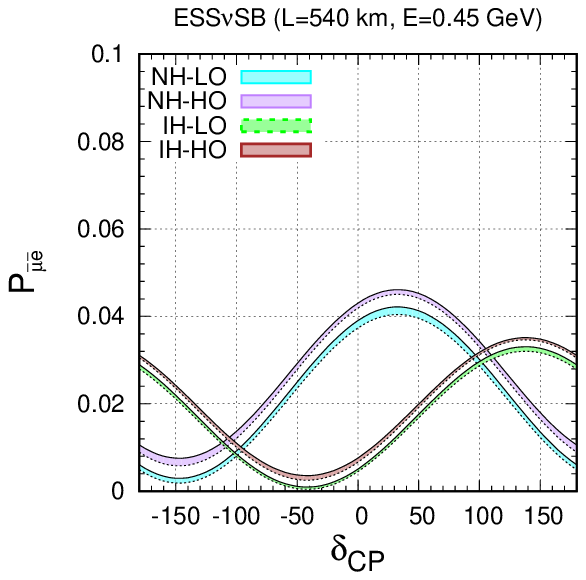}
	\end{tabular}
	\end{center}
\caption{Appearance probability for ESS$\nu$SB for different energies. The bands are obtained 
by varying $\theta_{23}$ in lower octant and upper octant. See text for details.}
\label{fig:prob-ESS}
\end{figure}

In Fig.\ref{fig:probPlots} we have plotted the appearance probabilities for 
\t2k baseline of 295 km (top row) and \nova baseline of 810 km  (bottom row) 
as a function of $\delta_{CP}$. The probability plot in this figure and the subsequent figures have been generated by exact numerical calculations using GLoBES. The energies are fixed at the peak energies of 
0.6 GeV and 2 GeV respectively in these plots. 
The left and right columns represent the neutrino and anti-neutrino oscillation probabilities respectively.  
Each plot in Fig.\ref{fig:probPlots} and Fig.\ref{fig:prob-ESS} 
comprises of four different hierarchy-octant bands NH-LO (cyan), NH-HO (purple), IH-LO (green), IH-HO (brown). Each LO (HO) band represents the \th23 variation in the range $39^\circ-42^\circ$ ($48^\circ-51^\circ$).   
These plots help to understand the various degeneracies occurring between 
hierarchy, octant and $\delta_{CP}$. 

With respect to \dcp there can be two kind of solutions : 
(i) those with wrong values of \dcp which can be seen by drawing 
a horizontal line through the curves and seeing the CP values 
at which this line intersects the two bands of opposite hierarchies and/or octant. 
(ii) Those with right values of \dcp which occur when two bands intersect each other. 
Thus we can have the following type of degenerate solutions  in addition 
to the true solution which can affect the hierarchy and octant sensitivity \footnote{Note that Right Hierarchy-Right Octant-Wrong $\delta_{CP}$ degeneracy is not a degeneracy for mass hierarchy and octant sensitivities and therefore it was not mentioned. Also, Right Hierarchy-Wrong Octant-Right $\delta_{CP}$ is not really a degeneracy. This can be understood by drawing a vertical line at a particular $\delta_{CP}$ in the $P_{\mu e}$ vs $\delta_{CP}$ plot, which leads us to the observation that the values of $P_{\mu e}$ will always be different for the opposite octants. If at all a solution encompassing LO and HO appears that will be due to the limitation in the measurement of $\theta_{23}$.} : 
\begin{itemize} 
\item Wrong Hierarchy - Right Octant - Right \dcp (WH-RO-R$\delta_{CP}$) 
\item Wrong Hierarchy - Right Octant - Wrong \dcp (WH-RO-W$\delta_{CP}$) 
\item Wrong Hierarchy - Wrong Octant - Right \dcp (WH-WO-R$\delta_{CP}$) 
\item Wrong Hierarchy - Wrong Octant - Wrong \dcp (WH-WO-W$\delta_{CP}$) 
\item Right Hierarchy - Wrong Octant - Wrong \dcp (RH-WO-W$\delta_{CP}$) 
\end{itemize} 

The plots in Fig.\ref{fig:probPlots} 
show that there are no degeneracies at the highest and lowest points 
of the probability bands. These correspond to NH-HO (NH-LO) and \dcp $\sim -90^\circ$ and  IH-LO (IH-HO)  and \dcp $\sim +90^\circ$ for neutrinos (anti-neutrinos).  
The rest of the combinations are not free from  hierarchy - octant - \dcp degeneracies \cite{Ghosh:2015ena}. 
\nova baseline being higher, shows 
a wider separation between opposite hierarchies than that of T2K. 
Also one can note that combination of neutrinos and anti-neutrinos can 
remove octant degeneracy. For instance NH-LO with \dcp $\sim -90^\circ$ is degenerate with IH-HO with \dcp in the same half plane and NH-HO in the opposite 
half plane of \dcp as can be seen by comparing the cyan, brown and 
purple bands for the neutrinos.
Thus with neutrinos one can get degenerate solutions
corresponding to WH-WO-R\dcp 
and RH-WO-W$\delta_{CP}$.   
However, if one considers anti-neutrinos then these degeneracies are not present. 
Thus combination of neutrino and anti-neutrino data can help in 
alleviating degenerate solutions in opposite octant 
\cite{Prakash:2013dua,Ghosh:2015ena,Agarwalla:2013ju,Machado:2013kya,Coloma:2014kca}.

The \ess set-up that we consider corresponds to that discussed in 
\cite{Baussan:2013zcy} with a baseline of 540 km -- between Lund 
and Garpenberg mine. 
The flux peaks around  0.24 GeV 
with significant flux around 0.35 GeV which is close to the second oscillation 
maxima. 
The probability has a sharper variation at the second oscillation maxima, 
with \dcp leading to a higher CP sensitivity. 
Primarily three energy bins with  mean energy 0.25 GeV (E1), 0.35 GeV (E2), 
0.45 GeV (E3) 
contribute significantly to the hierarchy sensitivity.  
Thus, in order to understand the degeneracies for the 
\ess baseline of 540 km we have plotted the appearance 
probability for three different energies in Fig.\ref{fig:prob-ESS} 
where the top, 
middle and bottom row corresponds to 0.25 GeV, 0.35 GeV and 0.45 GeV respectively. 
The behavior of $P_{\mu e}$ for 0.35 GeV is somewhat similar to 
that of  T2K and \nova in Fig.\ref{fig:probPlots}.
However, the variation with \dcp is sharper, octant bands are narrower 
and the different curves intersect each other more number of times indicating
presence of wrong hierarchy and/or wrong octant solutions at right $\delta_{CP}$.  
There is no degeneracy for NH-HO (NH-LO) at \dcp = $-90^\circ$
and IH-LO (IH-HO) at 
\dcp = $+90^\circ$ for neutrinos(anti-neutrinos) as in case of T2K and NO$\nu$A. 
The degeneracies for the energies 0.25 and 0.45 GeV are different than the above.  
From top left and bottom left plots in Fig.\ref{fig:prob-ESS}, 
we find that for neutrinos, IH does not suffer from 
mass hierarchy degeneracy in ($-180^\circ~ \le $ \dcp $\le -60^\circ~ $  and $130^\circ~ \le $ \dcp $\le 180^\circ~ $ ) 
but NH suffers from hierarchy degeneracy for  the whole range of $\delta_{CP}$.
For anti-neutrinos (top  and bottom right panel in Fig.\ref{fig:prob-ESS}) 
for $-30^\circ~ \ge $ \dcp $\ge ~ 70^\circ$ there is no degeneracy for NH while 
IH is degenerate  with NH throughout the full range of $\delta_{CP}$. 
Thus the degeneracies for NH and IH occur for different \dcp values 
and combination of neutrino and anti-neutrino runs can help in resolving these. 

The dependence of $P_{\mu e}$  on \dcp and 
$\theta_{23}$  can  be understood analytically by expressing the probability
$P_{\mu e}$  
as follows \cite{Agarwalla:2013ju}: 
\begin{eqnarray}
P_{\mu e}=(\beta_{1} - \beta_3) \sin^2\theta_{23} + 
\beta_{2} \sin 2\theta_{23} \cos(\Delta+\delta_{CP}) + \beta_{3}
\label{eq:pmuenew} 
\end{eqnarray}
where,
\begin{eqnarray}
\beta_{1} &=& \sin^2 2{\theta_{13}}
\frac{\sin^2\Delta(1 - \hat{A})}{(1 - \hat{A})^2},  \nonumber \\
\beta_{2} &=& \alpha\cos{\theta_{13}}\sin2{\theta_{12}}\sin2{\theta_{13}}\frac{\sin\Delta\hat{A}}{\hat{A}}
\frac{\sin\Delta(1-\hat{A})}{1-\hat{A}}, \nonumber \\
\beta_{3} &=& \alpha^2\sin^22{\theta_{12}}\cos^2{\theta_{13}}\frac{\sin^2\Delta\hat{A}}{\hat{A}^2}
\label{eq:betas}
\end{eqnarray}
\begin{table}[h!]
\begin{center}
\begin{tabular}{|c|c|c|c|c|}
\hline
 Baseline(L) & Peak Energy(E) &  $\beta_1$ & $\beta_2$  & $\beta_3$  \\
 \hline
   &   &  NH $~~~~~$ IH & NH $~~~~~$ IH  & NH $~~~~~$ IH  \\
\hline
295 km & 0.6 GeV &  0.094 $~~~~~$ 0.077 & 0.013 $~~~~~$ -0.011 & 0.002 $~~~~~$ 0.002 \\
810 km & 2.0 GeV &  0.095 $~~~~~$ 0.062 & 0.011 $~~~~~$ -0.009 & 0.001 $~~~~~$ 0.001 \\
540 km & 0.25 GeV & 0.015 $~~~~~$ 0.035 & 0.023 $~~~~~$ -0.035 & 0.034 $~~~~~$ 0.034\\
540 km & 0.35 GeV & 0.090 $~~~~~$ 0.071 & -0.039 $~~~~~$ 0.035 & 0.017 $~~~~~$ 0.017\\
\hline
\end{tabular}
\caption{$ \beta_1 $, $\beta_2$ \& $\beta_3$ values in Eq.\ref{eq:betas} for 295 km, 810 km, and 540 km baselines corresponding to T2K, \nova and \ess experiments.The energies correspond to the values where the flux peaks. For \ess we present the values for 
two representative energies.}
\label{tab:beta}
\end{center}
\end{table} 
The CP dependence of the probabilities can be understood from the following expression
\begin{equation}
P_{\mu e}(\delta_{CP})-P_{\mu e}(\delta_{CP}^{\prime}) = -2 \beta_{2} \sin 2\theta_{23} \sin \left( \Delta + \frac{\delta_{CP}+\delta^\prime_{CP}}{2} \right) \sin \left( \frac{\delta_{CP}-\delta^\prime_{CP}}{2} \right) 
\label{eq:delcpdiff} 
\end{equation}
for neutrino probabilities and normal hierarchy.
T2K and \nova are experiments close to first oscillations maxima 
corresponding to  $\Delta \approx \frac{\pi}{2}$.  Then, 
Eq.\ref{eq:delcpdiff} reduces to 
\begin{equation}
P_{\mu e}(\delta_{CP})-P_{\mu e}(\delta_{CP}^{\prime}) = -2 \beta_{2} \sin 2\theta_{23} \cos \left( \frac{\delta_{CP}+\delta^\prime_{CP}}{2} \right) \sin \left( \frac{\delta_{CP}-\delta^\prime_{CP}}{2} \right) 
\label{eq:delcpdiff-t2knova} 
\end{equation}  
For \ess as the bin 0.35 GeV is close to the second oscillations maxima, 
we can write $\Delta \approx \frac{3 \pi}{2}$. 
Hence, the equation \ref{eq:delcpdiff} is,  
\begin{equation}
P_{\mu e}(\delta_{CP})-P_{\mu e}(\delta_{CP}^{\prime}) = 2 \beta_{2} \sin 2\theta_{23} \cos \left( \frac{\delta_{CP}+\delta^\prime_{CP}}{2} \right) \sin \left( \frac{\delta_{CP}-\delta^\prime_{CP}}{2} \right) 
\label{eq:delcpdiff-ess2} 
\end{equation} 
The 0.25 GeV bin in \ess is closer to the third oscillations maxima, so $\Delta \approx \frac{5 \pi}{2}$. Hence, the governing equation for this energy is the Eq.\ref{eq:delcpdiff-t2knova}.
Although Eq.\ref{eq:delcpdiff-t2knova} and Eq.\ref{eq:delcpdiff-ess2}
have a relative $(-)$ sign, the shapes are similar for Fig.\ref{fig:probPlots} and the second row for Fig.\ref{fig:prob-ESS} because the $\beta_2 = 0.013$ for T2K, $\beta_2 = 0.011$ for \nova and $\beta_2 = -0.039$ for \ess.
Thus the negative sign in $\beta_2$ compensates for the relative negative signs between  the  two equations. 
The  sharper variation in the probabilities for \ess
can be attributed to the higher $|\beta_2|$ value of \ess 
as compared to T2K and \nova (
$|\beta_2|_{\rm{ESS}\nu\rm{SB}} \approx 3 |\beta_2|_{\rm{T2K(NOvA)}}$). 
To understand the shape of the $P_{\mu e}$ curve the slopes of the 
probability for various \dcp values should be understood, which is given by
\begin{equation}
S = \frac{d P_{\mu e}}{d \delta_{CP} } = - \beta_{2} \sin 2\theta_{23} \sin \left( \Delta + \delta_{CP} \right) 
\label{eq:slope} 
\end{equation} 
From Eq.\ref{eq:slope} we obtain that in Fig.\ref{fig:probPlots} the slope is positive from $-180^\circ < \delta_{CP} < -90^\circ$ and $90^\circ < \delta_{CP} < 180^\circ$, with the slope becoming zero at \dcp $ = -90^\circ$ and $90^\circ$ and positive from $-90^\circ < \delta_{CP} < 90^\circ$. The same explanation is also valid for 0.35 GeV probability for \ess but the slopes being higher in \ess due to greater $|\beta_2|$. 
Comparing the NH-LO(blue) and NH-HO(purple) bands Fig.\ref{fig:prob-ESS} and Fig.\ref{fig:probPlots} we can see that the variation of NH probabilities for \ess 0.25 GeV is more rapid compared to T2K and \nova but less rapid compared to 0.35 GeV of ESS$\nu$SB. This is because the $|\beta_2|$ for 0.25 GeV bin in \ess is greater compared to T2K and \nova and less in comparison with 0.35 GeV bin of ESS$\nu$SB. 
Similarly, the shapes for IH and  the anti-neutrino probabilities can be 
explained. 

The dependence of $P_{\mu e}$  on $\theta_{23}$  can also be understood
 from Eq.\ref{eq:pmuenew},\ref{eq:betas} and Tab.\ref{tab:beta}.
In case of \ess one can see from the plots in Fig.\ref{fig:prob-ESS} that the 
octant bands for 0.25 GeV and 0.45 GeV are very narrow 
indicating that the probabilities do not change significantly with
$\theta_{23}$. 
For the energy bin 0.35 GeV the octant bands are slightly 
wider as compared to the other two energies. 
The $\theta_{23}$ behavior is governed by  the first term of
Eq.\ref{eq:pmuenew}, which shows a  linear variation with $\theta_{23}$
with a slope of  ($\beta_1 - \beta_3$).
Over the  allowed range of $\theta_{23}$ 
($39^\circ - 51^\circ$), $\sin2\theta_{23} \sim  1$ and hence 
the second term of Eq.\ref{eq:pmuenew}
does not play a role in determining the dependence 
of $P_{\mu e}$ on $\theta_{23}$.
\begin{equation}
P_{\mu e}(\theta_{23})-P_{\mu e}(\theta_{23}^{\prime}) \approx (\beta_{1}-\beta_{3})(\sin^2\theta_{23}-\sin^2\theta_{23}^{\prime})
\label{eq:th23diff} 
\end{equation}
This  does not depend on \dcp which is corroborated by the probability
plots.  
The $\beta_{i}$s for the three different baselines  and the relevant 
energies are tabulated in
Tab.\ref{tab:beta} for both the hierarchies.
We can see that for the \ess baseline and 0.25 GeV energy
the  $\beta_1$ and $\beta_3$ are almost equal 
indicative of the fact that the probability does not vary much with
$\theta_{23}$. For 0.35 GeV energy $\beta_1 - \beta_3 \sim 0.73$ for NH 
and 0.54 for IH, hence the octant bands are comparatively 
wider. This also implies that the 
IH bands are slightly narrower as compared to the NH bands 
as can be seen from the figure. 
For 0.25 GeV and 0.45 GeV the  
octant degeneracy for the same hierarchy is seen to 
prevail over the full range of \dcp corresponding to  
RH-WO-R\dcp solutions.  
In addition RH-WO-W$\delta_{CP}$, WH-WO-W$\delta_{CP}$, WH-RO-W\dcp and WH-WO-R\dcp 
solutions are also seen to be present. 
The octant sensitivity of \ess comes mainly from the bin with mean energy
the 0.35 GeV. For this energy, the octant degeneracies are seen to occur close to $\pm 90^\circ$
in the same half plane giving WH-WO-R\dcp between NH-LO(cyan band) and IH-HO(brown band) for neutrinos and between NH-HO(purple band) and IH-LO(green band) for anti-neutrinos. 
For purposes of comparison we also give the $\beta_i$ values for T2K and NO$\nu$A. 
For these cases also the NH bands are slightly wider than the IH bands
as can be seen from the  Fig.\ref{fig:prob-ESS} and the values of ($\beta_1 - \beta_3$).

In comparison to the LBL experiments the atmospheric neutrinos 
in ICAL detector can travel through  larger baselines and encounter
resonance effects. At resonance, the probabilities can be better
described by the 
one mass scale dominance (OMSD) approximation rather than the 
$\alpha-s_{13}$ 
approximation in Eq.\ref{eq:pmue}. 
The relevant probabilities $P_{\mu \mu}$  and $P_{\mu e}$ 
in the OMSD approximation can be expressed as,  
\begin{equation}
{\mathrm{P^{m}_{\mu e}}} =
{\mathrm{
\sin^2 \theta_{23} \; \sin^2 2 \theta^m_{13} \; \sin^2\left[1.27 ~\dam ~\frac{L}{E} \right]
}}
\label{eq:pemumat}
\end{equation}
\bea
{\mathrm{P^{m}_{\mu \mu} }} &=&
{\mathrm{1 - \cos^2 \theta^m_{13} \; {\mathrm{\sin^2 2 \theta_{23}}}
\;
\sin^2\left[1.27 \;\left(\frac{\da + A + \dam}{2}\right) \;\frac{L}{E} \right]}}
\nonumber \\
&& ~-~
{\mathrm{
\sin^2 \theta^m_{13}\; {\mathrm{\sin^2 2 \theta_{23}}}
\;
\sin^2\left[1.27 \;\left(\frac{\da + A - \dam}{2}\right) \;\frac{L}{E}
\right]}}
\nonumber \\
&& ~-~
{\mathrm{\sin^4 \theta_{23}}} \;
{\mathrm { \sin^2 2\theta^m_{13}}} \;
\sin^2 \left[1.27\; \dam  \;\frac{L}{E} \right]
\eea

Due to matter effect the modified mass squared difference ${\mathrm{\dam}}$ and mixing angle ${\mathrm {\stsmallm}}$ are given by ,  
\bea
{\mathrm{\dam}} &=&
{\mathrm{
\sqrt{(\da \cos 2 \theta_{13} - A)^2 +
(\da \sin 2 \theta_{13})^2} }}
\nn \\
\nn \\
{\mathrm {\sin 2 \theta^m_{13} }}
&=& \frac{\mathrm{\da \sin 2 \theta_{13}}}
{{\mathrm{\sqrt{(\da \cos 2 \theta_{13} - A)^2 +(\da \sin 2 \theta_{13})^2} }}}
\label{eq:dm31}
\eea 
The Mikheyev-Smirnov-Wolfenstein(MSW) matter resonance~\cite{msw1,msw2,msw3} 
occurs when, $$\da \cos 2 \theta_{13} = A$$ 
The MSW resonance happens when $\da >0$ for neutrinos and $\da<0$ for anti-neutrinos.
Sine the resonance conditions are opposite for neutrinos and anti-neutrinos therefore the ability to distinguish between neutrinos and anti-neutrinos is crucial for mass hierarchy sensitivity. Hence, the ICAL detector which has charge sensitivity can help unfold the 
mass hierarchy. The $\sin^2\theta_{23}$ term in $P_{\mu e}$ and $\sin^4\theta_{23}$ term in $P_{\mu \mu}$ is responsible for the octant sensitivity of ICAL \cite{Brahmachari:2003bk}.
Detailed discussion on the octant dependence of the probability 
for atmospheric neutrinos can be found for instance in \cite{Chatterjee:2013qus}. 


\section{Experimental details}\label{sec:Experimental-details}

In this section we provide a brief description of the experiments 
used in our study --- the 
currently running long-baseline experiments T2K, NO$\nu$A and also 
the future proposed long-baseline experiment ESS$\nu$SB 
along with the atmospheric 
neutrino experiment ICAL@INO. 

T2K (Tokai to Kamioka) \cite{Abe:2017uxa} is a 295 km baseline experiment with 
the flux centered at the  neutrino energy around 0.6 GeV generated by
JPARC neutrino beam 
facility at power level higher than 300 kW. T2K has already collected $3.1 \times 10^{21}$ protons on target(POT) and is expected to collect a total of  $8 \times 10^{21}$ in 10 years. So, we have used a total POT of $8 \times 10^{21}$ in our simulation.
In order to minimize the experimental uncertainty a near and a far detector are used at an angle $2.5^\circ$ from the center of the neutrino beam. 
Using the shape of the $\hat{\rm{C}}$herencov rings, the Super Kamiokande detector 
(fiducial volume 22.5 kt) for T2K has the ability to distinguish between 
electron and muon events. In our analysis we have considered 4 years of
neutrino and 4 years of anti-neutrino runs for T2K. 

NO$\nu$A experiment \cite{Adamson:2017gxd} is also a long-baseline neutrino experiment with 
a baseline of 810 km between the source and the  detector.
 NO$\nu$A uses high intensity 400 kW 
NuMI beam at Fermi lab. In this experimental set up a relatively smaller, 
222 ton near detector and a bigger 15 kiloton far 
detector are placed at an 
off axis of $0.8^\circ$ from the NuMI beam with peak energy at 2 GeV.  
In both near and far positions liquid scintillator type detectors 
are used.  
\nova is currently running at 700 kW beam power corresponding to $7.3 \times 10^{20}$ POT  
yearly and has already collected $8.85 \times 10^{20}$ POT.   
\nova is planned to run in 3 years neutrino and 3 years of anti-neutrino mode. In our study we have used the re-optimized NO$\nu$A set up from 
\cite{Agarwalla:2012bv,patt} and have used the full projected run time. %

ESS$\nu$SB \cite{Baussan:2012cw,Baussan:2013zcy} is a 540 km baseline experiment where high intensity neutrino beam will be produced at the European Spallation Source (ESS) in Lund, Sweden. 
ESS$\nu$SB \cite{Wildner:2015yaa} will focus its research on the 
second oscillation maximum and plans to use a  
megaton water $\hat{\rm{C}}$herencov detector. 
To create a high intensity neutrino beam, they propose to use 
the linac facility of the European Spallation Source which will produce
2 GeV protons with an average beam
power of 5 MW and $27 \times 10^{23}$ POT.  
The \ess collaboration advocates 
an optimized set up with 2 years of neutrino and 8 years of anti-neutrino runs 
\cite{Baussan:2013zcy}.

$\bf{I}$ndia-based $\bf{N}$eutrino $\bf{O}$bservatory (INO) \cite{Kumar:2017sdq} is a proposal for observing atmospheric neutrinos in a magnetized $\bf{i}$ron $\bf{cal}$orimeter (ICAL) detector. This experiment will look for 
atmospheric $\nu_\mu$ and $\bar{\nu}_\mu$ in the GeV energy range.
It is proposed to  be built in the 
southern part of India under a mountain with 1 km overall rock coverage. 
It will house a 50 kt ICAL detector with 1.5 Tesla magnetic field. Because of the magnetic properties of the detector, ICAL can 
identify the polarity of the charged particles produced by the
$\bf{c}$harge $\bf{c}$urrent (CC) interaction of neutrinos with the detector. 
This gives it the ability to differentiate 
between neutrinos and anti-neutrinos by identifying the charge of the daughter
particles using Resistive Plate Chambers (RPC) as an active 
detector component. 

\section{Simulation details}\label{sec:Simulation details}

In this section we present the details of our simulation procedure 
for the LBL experiments T2K, \nova and \ess and the 
atmospheric neutrino experiment ICAL@INO. We also 
discuss how the combined analysis of the various experiments are performed.  

The simulation for the 
long-baseline accelerator experiments \t2k , \nova \& \ess are done
using the General Long Baseline Experiment Simulator (GLoBES) package \cite{Huber:2004ka, Huber:2007ji}.  
In this, the capability of an experiment to determine an oscillation parameter is obtained by a  $\chi^2$-analysis using frequentist approach. 
The total $\chi^2_{{\rm tot}}$ is composed of $\chi^2_{{\rm stat}}$ and $\chi^2_{{\rm pull}}$ and is  given by the following relation
\begin{eqnarray}
\chi^2_{{\rm tot}} = \underset{\xi, \Omega}{\mathrm{min}} \left\lbrace \chi^2_{{\rm stat}}(\Omega,\xi) + \chi^2_{{\rm pull}}(\xi) \right\rbrace 
 \label{chi-tot}
\end{eqnarray}
where $\Omega = \{\theta_{12}, \theta_{13}, \theta_{23}, \Delta m^2_{21}, \Delta m^2_{31}, \delta_{CP}\}$ represents the oscillation parameters, 
$\chi^2_{{\rm stat}}$ denotes the poissonian $\chi^2$ function and $\chi^2_{{\rm pull}}$ consists of the systematic uncertainties incorporated in terms of pull variables $(\xi)$. 
The ``pull" variables considered in our analysis are signal 
normalization error, background normalization error,
energy calibration error on signal \& background (tilt).
In the ``pull" method a penalty term is added in terms of the ``pull" variables which is given by $\chi^2_{{\rm pulls}} = \sum\limits_{r=1}^{r=4} \xi^2_{r}$ in order to account for the systematics errors stated above. 
The poissonian $\chi^2_{{\rm stat}}$ is given by 
\begin{eqnarray}
 \chi^2_{{\rm stat}}(\Omega,\xi)= 2 \sum\limits_{i} \left\lbrace  \tilde{N_i}^{{\rm test}}-N_i^{{\rm true}}+N_i^{{\rm true}} \ln \frac{N_i^{{\rm true}}}{\tilde{N_i}^{{\rm test}}} \right\rbrace  .
 \label{chi-stat}
\end{eqnarray} 
The number of events predicted by the theoretical model over a range of oscillation parameters $\Omega$ in the 
$i^{th}$ bin and is given by 
\begin{eqnarray}
\tilde{N_i}^{{\rm test}}(\Omega, \xi) = \sum\limits_{k = s, b} {N_i}^{k}(\Omega) \left[  1 + c_{i}^{(k) norm} \xi^{(k) norm} + c_{i}^{(k) tilt} \xi^{(k) tilt} \frac{E_i - \bar{E}}{E_{max} - E_{min}} \right] 
\end{eqnarray}
where, $k = s(b)$ denotes signal(background) and $c_i^{norm}$(${c_i}^{tilt}$) represents alteration in the number of events by the
modification of the ``pull" variable $\xi^{norm}$(${ \xi}^{tilt}$). $E_i$ is the mean reconstructed energy of the $i^{th}$ bin, and $\bar{E} = ({E_{max} +E_{min}})/{2}$ is the mean 
energy over this range with 
$E_{min}$ and $E_{max}$ denoting the  maximum and minimum energy.
The systematic errors on the signal and
background normalizations are shown in Tab.\ref{tab:systLBL} \footnote{Note that, we have used statistical errors as dominant for $\nu_{\mu}$ and $\bar{\nu_{\mu}}$ in case of T2K\cite{Kato_2008}. Therefore, the errors for the disappearance channels are kept small. }.
 $N_i^{{\rm true}}$ in Eq.\ref{chi-stat} is given by the sum of simulated signal and background events $N_i^{{\rm true}} = N_i^s + N_i^b$. \\
The background channels influencing detection of neutrinos is dependent on the type of detector used. 
The background channels which contribute for  the water 
$\hat{\rm{C}}$herenkov detectors for 
\t2k and \ess are the  Charged Current(CC) non-Quasielastic(QE) background, intrinsic beam background, neutral current background and mis-identification error. While the main background channels affecting the scintillator detector in \nova are CC non-QE, intrinsic beam background, neutral current background. The systematic uncertainties in signal and background normalizations in various channels are summarized in Tab.\ref{tab:systLBL}. Besides these, the energy calibration errors are also incorporated in the analysis in terms of ``tilt" errors. The signal (background) ``tilt" errors that have been included are 1\%(5\%) for T2K, 0.1\%(0.1\%) for \nova and 0.1\%(0.1\%) for \ess. 

\begin{table}[H]
\begin{center}
\begin{tabular}{|l|c|c|r|}
\hline
Channel & \t2k & \nova & \ess \\
\hline \hline
$\nu_e$ appearance &  2\% (5\%) & 5\% (10\%)  & 5\% (10\%) \\
\hline
$\bar{\nu}_e$ appearance & 2\% (5\%)  &  5\% (10\%) & 5\% (10\%) \\
\hline
$\nu_\mu$ disappearance &  0.1\% (0.1\%) & 2.5\% (10\%)  &  5\% (10\%) \\
\hline
$\bar{\nu}_\mu$ disappearance & 0.1\% (0.1\%)  &  2.5\% (10\%) & 5\% (10\%) \\
\hline 
\end{tabular}
\caption{The signal (background) normalization errors for \t2k , \nova and \ess . }
\label{tab:systLBL}  
\end{center}
\end{table}

%

ICAL is a 50 kt detector which aims to detect $\nu_\mu$ and $\bar{\nu}_\mu$ along with hadron produced in the detector. Atmospheric flux consist of $\nu_\mu$ ($\bar{\nu}_\mu$) and $\nu_e$($\bar{\nu}_e$) both of which will contribute to the number of events observed in the ICAL detector. The events observed in the detector can be expressed as:

\begin{equation}
\frac{d^2N}{d\Psi_\mu} = (t~n_d) \times \int^{}_{} d\Psi_\nu ~d\Phi_\mu~  \biggl[ P_{\mu\mu} \frac{d^3\Phi_\mu}{d\Psi_\nu ~d\Phi_\nu} + P_{e\mu} \frac{d^3\Phi_e}{d\Psi_\nu ~d\Phi_\nu}   \biggr] \frac{d\sigma_\mu (E_\nu)}{d\Psi_\mu}
\end{equation} 
where, $d\Psi_\alpha = dE_\alpha ~ d\cos\theta_\alpha$, $n_d$ is the number of nucleon target in the detector, \emph{t} is the experiment run time, $\Phi_\mu$ and $\Phi_e$ are the initial flux of muon and electron respectively, $\sigma_\mu$ is the differential neutrino interaction cross section. $P_{\mu\mu}$ and $P_{e\mu}$ are the muon survival and appearance probabilities. To reduce the Monte Carlo fluctuations, firstly 1000 years of unoscillated data is generated with Nuance \cite{Casper:2002sd} neutrino generator using Honda neutrino flux and the interaction cross section and the ICAL detector geometry. 
Each event is then multiplied with the oscillation probability depending on 
the neutrino energy and path length. Oscillation probability in matter is 
calculated solving differential neutrino propagation equation in matter using the PREM model for the density profile of the Earth \cite{Dziewonski:1981xy}. 
These events are smeared on a bin by bin basis using detector resolutions and 
efficiencies \cite{Chatterjee:2014vta, Devi:2013wxa} to get realistic simulation of the ICAL detector. Later in our analysis we have scaled down to 10 years of ICAL data.  
For ICAL@INO both ``data" and theory events are simulated in the same way. 
In our analysis, we have used the resolution and the 
efficiency obtained for the central part of the ICAL detector 
\cite{Chatterjee:2014vta} 
using GEANT4 \cite{AGOSTINELLI2003250, ALLISON2016186, 1610988} based 
simulation for the whole detector. 
The typical muon energy resolution in the GeV energy range is $\sim 10\%$ and angular resolution is $\sim 1^\circ$ and charge identification efficiency is greater than $\sim 95\%$ in the relevant energy range\cite{Thakore:2013xqa}. As ICAL has very good energy and angular resolution for the muons produced in the detector, so the events spectrum is categorized 
in terms of measured muon energy ($E_\mu$) and reconstructed muon angle ($\cos\theta_\mu$). This binning scheme is referred to as $\bf{2D}$ binning scheme in our analysis. Hadrons are also produced along with muons in the CC interaction with the detector. The hadron resolution is $\sim 85\%$ at $1$ GeV and $\sim 36\%$ at $15$ GeV. It is also possible to extract the hadron energy in event by event basis in CC interaction as $E^\prime_h = E_\nu - E_\mu$. Now the inelasticity parameter $y = \frac{E^\prime_h}{E_\nu}$ can be used as an independent parameter with every CC event in the previously mentioned 2D scheme. Analysis with these three independent parameters is referred to as the $\bf{3D}$ binning scheme. 
Improvement of the sensitivities while using the 3D binning scheme over the 2D
scheme has been shown in \cite{Devi:2014yaa}.  
The binning scheme is summarized in Tab.\ref{table:3D_Binning_Scheme}. While generating the data, oscillation parameters are used at their true values whereas the test events are generated using the $3\sigma$ range as mentioned in the Tab.\ref{tab:oscParam}. 
 
\begin{table}[H]
\centering
\begin{tabular}{|c|c|c|}
\hline
Reconstructed variable  & [ Range ] (bin width) & Total bins \\
\hline \hline
 & [1 : 4] (0.5) & 6 \\
 $E^{obs}_{\mu}$ (GeV) & [4 : 7] (1) & 3 \\
 & [7 : 11] (4) & 1 \\
 \hline
 & [-1.0 : -0.4] (0.05) & 12 \\
 $\cos\theta_{\mu}$ & [-0.4 : 0] (0.1) & 4 \\
 & [0.0 : 1.0] (0.2) & 5\\
 \hline
 & [0.0 : 2.0] (1) & 2\\
 $E'_{\rm{had}}$ (GeV) & [2.0 : 4.0] (2) & 1 \\
 & [4.0 : 11.0] (11) & 1 \\
 \hline
\end{tabular}
\caption{The binning scheme used in 2D ($E_\mu$, $\cos\theta_\mu$) and 3D ($E_\mu$, $\cos\theta_\mu$, $E'_{\rm{had}}$) analysis. 
}
\label{table:3D_Binning_Scheme}
\end{table}

As ICAL is not sensitive to \dcp\cite{Gandhi:2007td}, we have not marginalized over test \dcp 
and fixed the value at $0^\circ$ while generating the test events. To get the ICAL@INO sensitivity we perform a $\chi^2$ analysis where with $
chi^2$ defined as:

\begin{eqnarray} 
\chi^2_{\rm{INO}} & = & {\stackrel{\hbox{min}}{\displaystyle \xi_l^{\pm}}} 
\sum^{N_{E^{obs}_{\mu}}}_{i=1}\sum^{N_{\cos\theta^{obs}_{\mu}}}_{j=1}
\sum^{N_{E'^{obs}_{had}}}_{k=1} 2\left[ \left(T^{+}_{ijk}
- D^{+}_{ijk} \right) - D^{+}_{ijk} \ln \left( 
\frac{T^{+}_{ijk}}{D^{+}_{ijk}} \right) \right] + \nonumber \\
& & 2\left[\left(T^{-}_{ijk} - D^{-}_{ijk}\right) - D^{-}_{ijk}
\ln\left(\frac{T^{-}_{ijk}}{D^{-}_{ijk}}\right)\right] + 
\sum^{5}_{l^{+}=1} \xi^{2}_{l^{+}} + \sum^{5}_{l^{-}=1}
\xi^{2}_{l^{-}} ~.
\label{chisq-11p}
\end{eqnarray}
Where $i,j,k$ sums over muon energy, muon angle and hadron energy bins respectively. $T_{ijk}, ~D_{i,j,k}$ refers the predicted (theory) and observed (data) events respectively in $i, j, k$ bin. The $\pm$ sign in theory or observed events refer to the $\mu^{\pm}$ events coming from $\nu^{\pm}_\mu$ interactions in the detector. The number of expected events with systematic errors in each bin are given by:  
\begin{equation}
 T^{+}_{ijk} = T^{0+}_{ijk}\left(1+\sum^{5}_{l^{+}=1} 
\pi^{l^{+}}_{ijk}\xi_{l^{+}}\right)~;~~
T^{-}_{ijk} = T^{0-}_{ijk}\left(1+\sum^{5}_{l^{-}=1}
\pi^{l^{-}}_{ijk}\xi_{l^{-}}\right)~.
\label{td-pi6xi6}
\end{equation}
where $T^{0\pm}_{ijk}$ refers  to 
the number of theory events without systematic errors in a particular bin $i,j,k$. The systematic uncertainties considered in the analysis using pull method are \cite{PhysRevD.70.033010}:  
\begin{itemize}
\item $\pi_1=20$\% flux normalization error
\item $\pi_2=10$\% cross section error
\item $\pi_3=5$\% tilt error
\item $\pi_4=5$\% zenith angle error
\item $\pi_5=5$\% overall systematics 
\end{itemize}
To incorporate the ``tilt" error, the predicted neutrino fluxes is modified using:
\begin{equation}
\Phi_\delta(E) = \Phi_0(E)\left( \frac{E}{E_0}\right) ^\delta \simeq \Phi_0(E)\left( 1+\delta\ln\frac{E}{E_0}\right) ,
\end{equation}
where, $E_0$ is 2~GeV, and $\delta$ is the 1$\sigma$ systematic ``tilt" error (5\%). Flux uncertainty is 
included as $\Phi_\delta(E)-\Phi_0(E)$.

\begin{table}[H]
\begin{center}
\begin{tabular}{|c|c|c|}
\hline
Oscillation parameters & True value & Test range \\
\hline \hline
$\sin^2 2\theta_{13}$ & 0.085  &  fixed \\
\hline
$\sin^2 \theta_{12}$ & 0.304  &  fixed \\
\hline
$\theta_{23}$ & $42^\circ$ (LO), $48^\circ$ (HO)  & $39^\circ : 51^\circ$ \\
\hline
$\Delta m^2_{21}$ (eV$^2$) & $7.4\times 10^{-5}$ & fixed \\
\hline 
$\Delta m^2_{31}$ (eV$^2$) & $2.5\times 10^{-3}$ & $(2.35 : 2.65)\times 10^{-3}$  \\
\hline
$\delta_{\rm{CP}}$ (LBL) & $-180^\circ : 180^\circ$ & $-180^\circ : 180^\circ$ \\
$\delta_{\rm{CP}}$ (INO) & $-180^\circ : 180^\circ$ & $0^\circ$(fixed) \\
\hline
\end{tabular}
\caption{The true and test values of the oscillation parameters used in our analysis. }
\label{tab:oscParam}  
\end{center}
\end{table}

For performing the statistical analysis the observed events or data 
are generated using the 
true values listed in Tab.\ref{tab:oscParam}.
The  predicted or test  
events are simulated varying  the parameters 
 $|\Delta m^2_{31}|$, $\sin^2\theta_{23}$ in their $3\sigma$ 
ranges presented  in Tab.\ref{tab:oscParam}. 
The values of $\theta_{12}$, $\sin^2\theta_{13}$ and $\Delta m^2_{21}$ are held fixed 
to their best-fit values while calculating the test events. 
For LBL experiments \dcp is varied over $0^\circ - 360^\circ$ while 
generating the test events. 
For calculating $\chi^2$  corresponding to hierarchy or 
octant sensitivity for a particular experiment 
marginalization is done over the oscillation parameters which are varied in the 3$\sigma$ range obtained from the global analysis of the current data.  If two or more LBL experiments are combined then the 
$\chi^2$ of each experiment are added in the test plane 
and then the combined  $\chi^2$ is marginalized over. 
Since ICAL@INO is insensitive to $\delta_{CP}$, the $\delta_{CP}$(test)
is  kept fixed for ICAL@INO analysis to save computation time. 
While calculating the combined $\chi^2$ for 
LBL and ICAL@INO, 
the marginalization over test-\dcp is first performed for LBL 
experiments and then the marginalized $\chi^2$ is added with the ICAL@INO $\chi^2$.
This $\chi^2$ is further marginalized over the oscillation parameters 
$|\Delta m^2_{31}|$,
$\sin^2\theta_{23}$ as follows : 
\begin{equation}
 \chi^2_{tot} ~=~ \stackrel{\hbox{Min}}{\hbox{\small{$\theta_{23}$,$|\Delta m^2_{31}|$}}}\biggl[ \chi^2_{INO} +  \stackrel{\hbox{Min}}{\hbox{\small{$\delta_{CP}$}}}  \chi^2_{LBL} \biggr].
 \label{eq:ch-lbl_ino}
\end{equation}


\section{Results and Discussions}\label{sec:Results and Discussions}

\subsection{Mass hierarchy sensitivity}


The mass hierarchy sensitivity is calculated by taking a true set of
parameters assuming NH(IH) as true hierarchy and is compared against the
test parameters assuming the opposite hierarchy IH(NH).
While calculating the hierarchy sensitivity, marginalization is done over
$\theta_{23}$, $|\Delta m^2_{31}|$ and \dcp in the range depicted in 
Tab.\ref{tab:oscParam} in the test events (unless otherwise mentioned) while
$\theta_{12}$, $\theta_{13}$ and $\Delta m^2_{21}$ are kept fixed at their
best-fit values.

\begin{figure}[h!]
\begin{center}
        \begin{tabular}{clr}
        \hspace{-3cm}
	\includegraphics[width=0.55\textwidth]{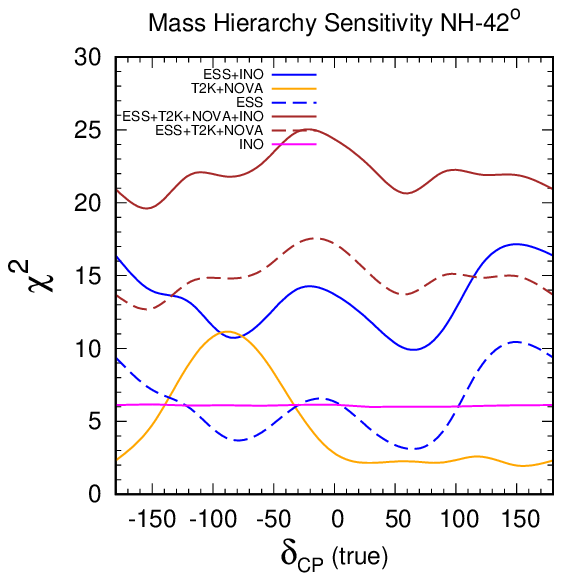}
	\hspace{-3cm}
        \includegraphics[width=0.55\textwidth]{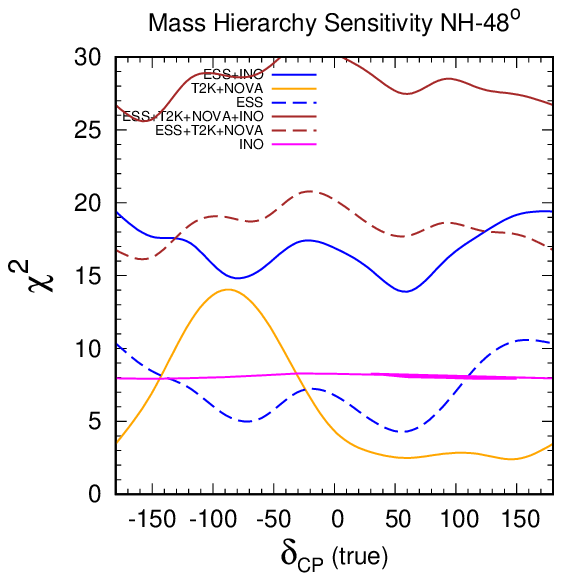}\\
        \hspace{-3cm}       
        \includegraphics[width=0.55\textwidth]{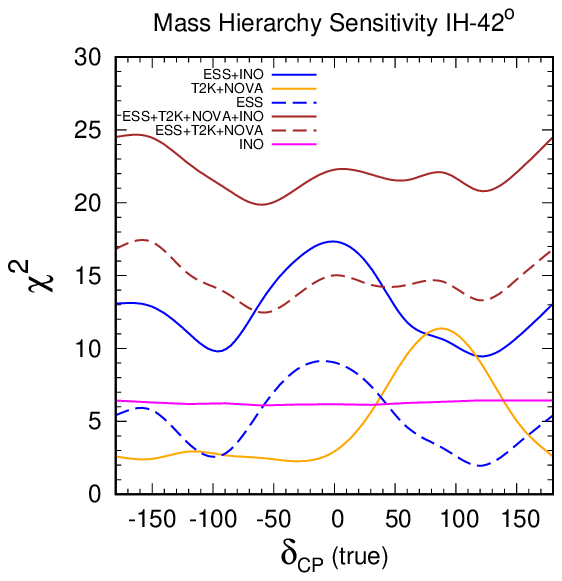}
	\hspace{-3cm}
        \includegraphics[width=0.55\textwidth]{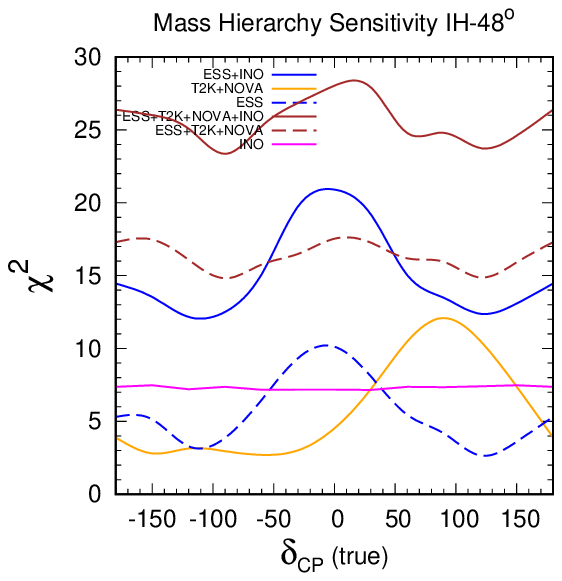}
\end{tabular}
\end{center}
\caption{Mass hierarchy sensitivity vs \dcp (true) for ICAL@INO-3D \ess T2K \nova for four hierarchy-octant combinations starting from top right, clockwise in the order NH-LO, NH-HO, IH-HO and IH-LO. Each figure consists of six different experimental combinations which are represented as ICAL@INO (magenta solid curve), \ess (blue dashed curve), T2K + \nova (orange solid curve), \ess + ICAL@INO (blue solid curve), \ess + T2K + \nova (brown dashed curve) and \ess + T2K + \nova + ICAL@INO (brown solid curve). }
\label{fig:MH-3D}
\end{figure}

In Fig.\ref{fig:MH-3D} we present the hierarchy
sensitivities as a function of true \dcp for the experiments \ess (2 years neutrino + 8 years anti-neutrino),
T2K + \nova (3 years neutrino + 3 years anti-neutrino),
INO-3D, individually and combined with each other
for four hierarchy-octant combinations. These are NH-LO, NH-HO, IH-LO and IH-HO.
The representative true values of $\theta_{23}$ for LO and HO are chosen as $42^\circ$ and
$48^\circ$ respectively.

The blue dashed lines in the plots represent the hierarchy sensitivity of
ESS$\nu$SB.
It is seen that for all the four cases
hierarchy sensitivity of \ess 
is more for CP conserving values ($0^\circ$, $\pm 180^\circ$)
than for CP violating values ($\pm 90^\circ$).
Overall, for NH and $\theta_{23} = 42^\circ$, \ess can have close to 
$2\sigma$ hierarchy sensitivity for all values of \dcp reaching
upto $3\sigma$ for \dcp = $\pm 180^\circ$.
The CP dependence of hierarchy sensitivity for NH-HO
is similar to that of NH-LO as can be seen from the right panel of the top row.
However, the sensitivity is slightly higher because of the higher octant.
The panels in the second row show the hierarchy sensitivity for IH-LO and IH-HO.
In this case \ess attains highest hierarchy sensitivity ($\sim 3 \sigma$) 
for \dcp$=0^\circ$ but sensitivity is $< 2 \sigma$ for $\delta_{CP} = \pm 90^\circ$. 

\begin{figure}[h!]
\begin{center}
        \begin{tabular}{clr}
        \hspace{-3cm}
        \includegraphics[width=0.55\textwidth]{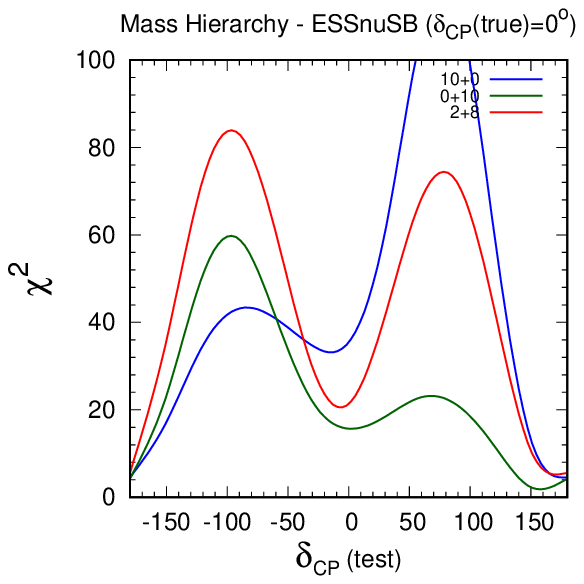}
	\hspace{-3cm}
	\includegraphics[width=0.55\textwidth]{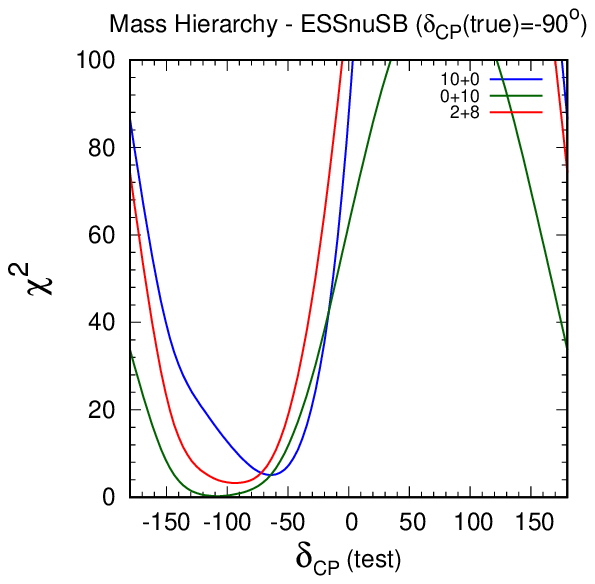}
	\hspace{-3cm}
        \includegraphics[width=0.55\textwidth]{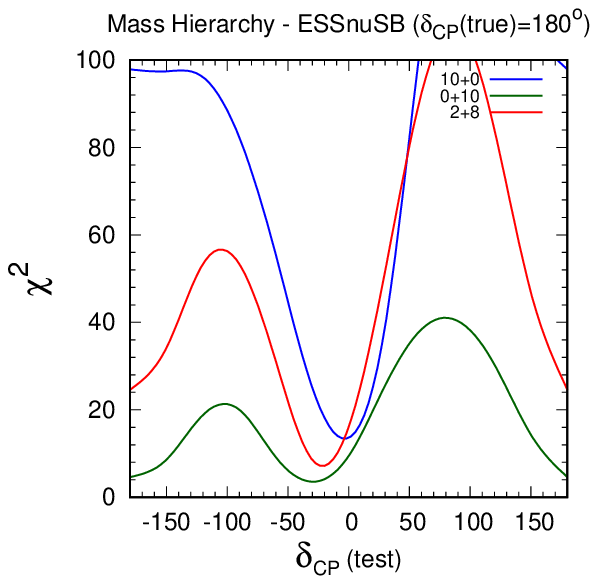}\\
         \hspace{-3cm}
         \includegraphics[width=0.55\textwidth]{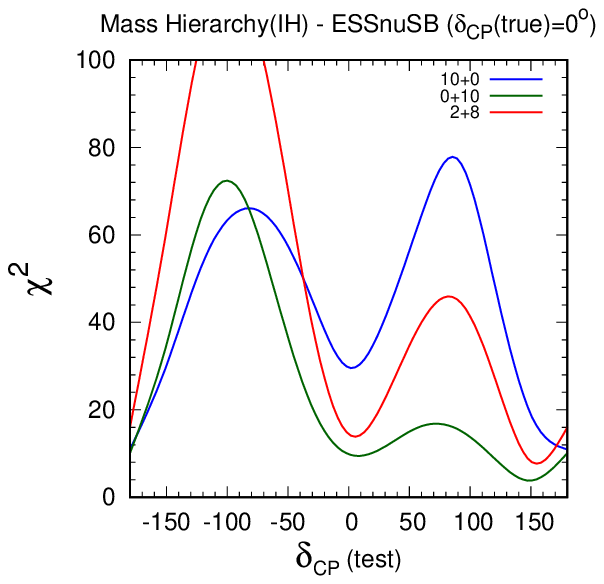}
	\hspace{-3cm}
	\includegraphics[width=0.55\textwidth]{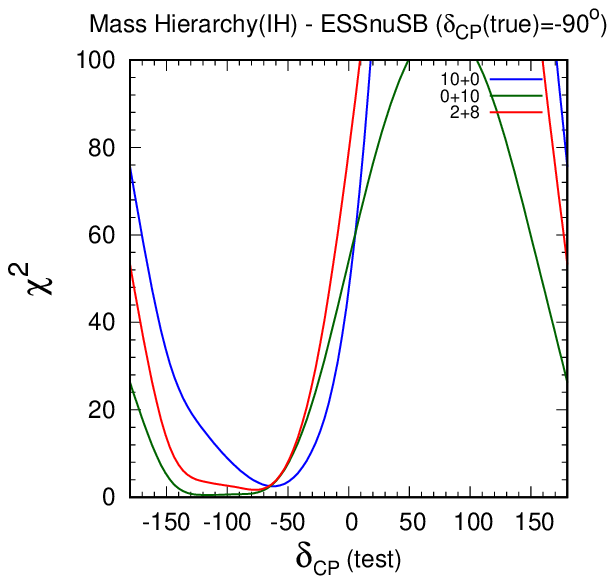}
	\hspace{-3cm}
        \includegraphics[width=0.55\textwidth]{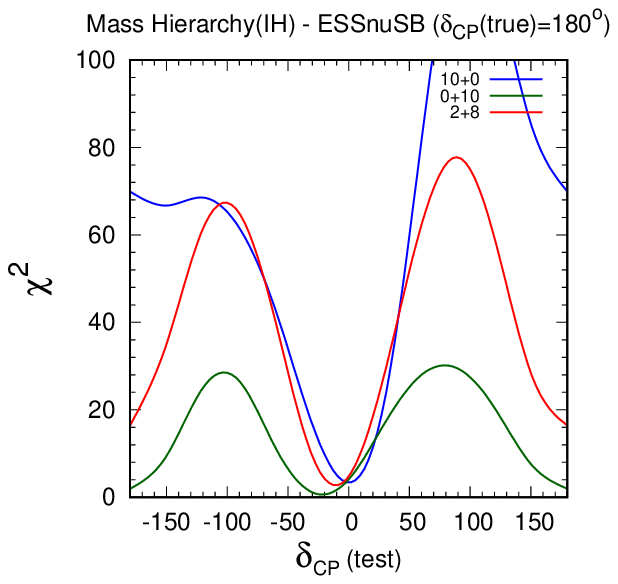}
\end{tabular}
\end{center}
\caption{Mass hierarchy sensitivity vs \dcp (test) for \ess with true \dcp fixed. The top row represents true NH white the bottom row represents true IH. The true octant is fixed as lower octant for all the plots. Each row has plots for three different \dcp (true), which are $0^\circ$,$-90^\circ$ and $180^\circ$ from left to right in the given order. Each figure consists of three different runtime combinations of neutrino and anti-neutrino runs. The blue, green and red curves denote the runtime combinations (in years) of $10\nu+0\bar{\nu}$, $0\nu+10\bar{\nu}$ and $2\nu+8\bar{\nu}$ respectively.}
\label{fig:mass-ess}
\end{figure}

In order to understand the behavior of the hierarchy sensitivity $\chi^2$ with
\dcp in Fig.\ref{fig:mass-ess} we plot the hierarchy $\chi^2$ as a function of
test-\dcp for only neutrino, only anti-neutrino and the mixed runs of 
10$\nu$+0$\bar{\nu}$ years, 0$\nu$+10$\bar{\nu}$ years 
and 2$\nu$+8$\bar{\nu}$ years respectively for the \ess experiment. 
One can see that for \dcp= $0^\circ$ the minimum for neutrinos come at
$\pm 180^\circ$ while that of anti-neutrinos occur at $150^\circ$.
Therefore, the position of the overall minimum combining neutrinos and anti-neutrinos is
at \dcp $= \pm 180^\circ$.
For \dcp(true) $= 180^\circ$, the neutrino minimum is seen to be at \dcp(test) $=0^\circ$
whereas the anti-neutrino minima at
\dcp(test) $\approx -30^\circ$  and $\pm 180^\circ$.
The overall minimum of 2$\nu$+8$\bar{\nu}$ years runtime come at \dcp $= -20^\circ$.
In comparison the figure drawn for true \dcp $= -90^\circ$  (the second panel in the first row) shows
that the minimum for only neutrino run comes around $-60^\circ$ whereas
that for only anti-neutrino comes around $-110^\circ$. Whereas the combined minimum occurs close to $-90^\circ$.
Thus for \dcp $=- 90^\circ$ the wrong hierarchy
minima comes at the same CP value as compared to \dcp(true) $=0^\circ$, $\pm 180^\circ$ where the wrong hierarchy minima are farther from the true value and hence the tension is enhanced.
Similar feature is observed in the IH-LO curves (in the lower panel) also. For \dcp(true) $= -90^\circ$
the minima for only neutrino, only anti-neutrino and the combined
runs occur in the same half plane as \dcp $= -90^\circ$ whereas for
true \dcp $=0^\circ$ the corresponding minima occurs at $\pm 180^\circ$
and  $150^\circ$.
Similarly for \dcp $= 180^\circ$, the minimum for only neutrino run
is at $0^\circ$, for only anti-neutrino run is at $-30^\circ$ and the combined
run is close to $-10^\circ$.
Thus for \dcp $= -90^\circ$ the presence of WH-R\dcp degeneracy gives a lower sensitivity as compared to the CP conserving values, where the wrong hierarchy minima come at different wrong CP values for neutrino and anti-neutrino which enhances the tension and hence the $\chi^2$.
It is seen that the only neutrino run of \ess can give a better hierarchy
sensitivity because of statistics than the combined runs. However,
the combined neutrino and anti-neutrino run is expected to give
a higher CP sensitivity \cite{Baussan:2013zcy}.

The magenta curves  in Fig.\ref{fig:MH-3D} represents the hierarchy sensitivity of ICAL@INO.
Mass hierarchy sensitivity of ICAL@INO is independent of \dcp because of the
sub-dominant effect of \dcp  in survival probability and also due to
smearing over directions \cite{Ghosh:2013zna,Ghosh:2014dba}.
Hence, when ICAL@INO is added to other long-baseline accelerator experiments a
constant increase in the sensitivity is observed.
This helps to get reasonable sensitivity in the degenerate region.
This is reflected by the blue solid curves which demonstrate
the hierarchy
sensitivity of ESS$\nu$SB + ICAL@INO.
Since ICAL@INO has no dependence on \dcp the combined curve follows
the \ess curve.
The combination of \ess + ICAL@INO can give $3\sigma$ hierarchy sensitivity over the
whole range of \dcp reaching  $4\sigma$ for \dcp = $\pm 180^\circ$
for NH-LO.  
For NH-HO, the combined sensitivity is more than $3.5\sigma$ for all values
of \dcp crossing $4\sigma$ for $\pm 180^\circ$.
For IH-LO, as can be seen from the second row first column of the Fig.\ref{fig:MH-3D}
the combined sensitivity of \ess+ ICAL@INO is more than $\sim 3\sigma$ for
all values of \dcp and more than $4\sigma$ for \dcp $=0^\circ$.
For IH-HO, the sensitivity at \dcp $=0^\circ$ can reach $\sim 4.5 \sigma$.

The yellow curves in the different panels of
Fig.\ref{fig:MH-3D}  represent the hierarchy sensitivity for T2K + NO$\nu$A.
We see that for NH-LO/HO highest sensitivity occurs for \dcp $\sim -90^\circ$ and the
lower half plane (LHP, $-180^\circ < $\dcp $< 0^\circ$) is seen to be favorable for hierarchy.
This can be understood from the
probability figures for T2K and \nova in Fig.\ref{fig:probPlots}. 
For instance 
NH-LO (the cyan band) and \dcp $\sim -90^\circ$
does not show any degeneracy for anti-neutrinos. 
The degeneracy with IH-HO present for neutrinos 
can be resolved when neutrino and anti-neutrino are combined. 
Thus the LHP is conducive for hierarchy determination \cite{Prakash:2012az}.
For true NH-HO, neutrino has no degeneracy for \dcp = $-90^\circ$, 
whereas the WH-WO-R\dcp degeneracy observed in the anti-neutrino data
(purple and green bands in the probability Fig.\ref{fig:prob-ESS} )
 can be resolved when combined with neutrino data.
Thus the LHP is favorable for hierarchy. 
For true IH-LO/HO the upper half plane(UHP, $0^\circ <$ \dcp  $<180^\circ$) is favorable for hierarchy, since there is no degeneracy for neutrinos (anti-neutrinos) 
for \dcp in the UHP. 

When T2K + \nova $\chi^2$ is added to \ess and marginalized, the CP dependence
of the hierarchy sensitivity is governed by all the three experiments.
The resultant curve shows highest sensitivity for \dcp $=0^\circ$ reaching $4\sigma$
level for both NH-LO and NH-HO.
For IH, $4\sigma$ sensitivity is reached for HO for all values of \dcp
and for \dcp $= \pm 180^\circ$ for LO.
Note that the hierarchy sensitivity for \ess+ ICAL@INO for certain values of \dcp
could be higher than that of T2K + \nova+ ESS$\nu$SB.

The brown solid curve represents the combined hierarchy sensitivity of
\ess+ T2K + \nova+ ICAL@INO and it shows sensitivity reach of $\chi^2 = 25$ for all values of \dcp
NH-HO. For NH-LO $5\sigma$ sensitivity is reached for \dcp $=0^\circ$, while for IH-HO \dcp $=0^\circ$, $\pm 180^\circ$. For IH-LO, the sensitivity stays slightly higher than $4.4\sigma$ for all values of $\delta_{CP}$.
We have also observed a synergy between ICAL@INO and the accelerator based long-baseline experiments owing to $|\Delta m^2_{31}|$ tension. This tension happens because atmospheric data slightly prefers lower $|\Delta m^2_{31}|$  values while the accelerator data supports the true value.

Note that, the above analyses are done using 
the $\chi^2$ defined in Eq.\ref{eq:ch-lbl_ino} where we have not 
included any priors. 
We have also checked the effect of priors over $\delta_{CP}$, $\Delta m^2_{31}$ and $\theta_{23}$ taking their $3\sigma$ ranges from the global analysis of the current data. In this case, we obtain up to 9\% increment in the sensitivity with prior on $\delta_{CP}$, while priors over $\Delta m^2_{31}$ and $\theta_{23}$ does not have any significant effect on the sensitivities. Hence, our results are
more conservative.

\begin{figure}[h!]
\begin{center}
        \begin{tabular}{clr}
        \hspace{-3cm}
        \includegraphics[width=0.55\textwidth]{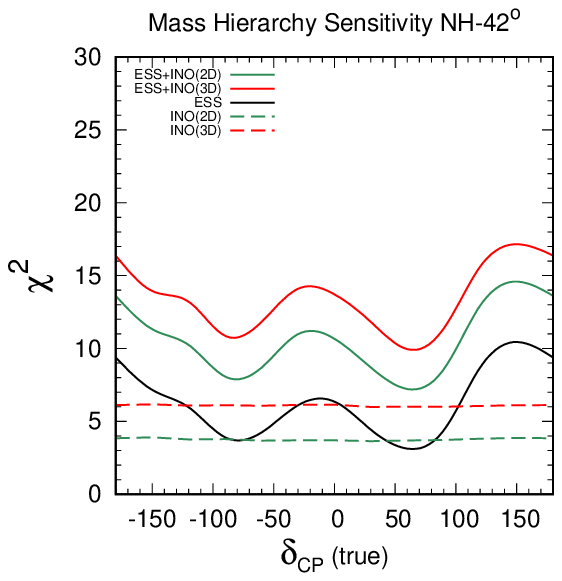}
	\hspace{-3cm}
        \includegraphics[width=0.55\textwidth]{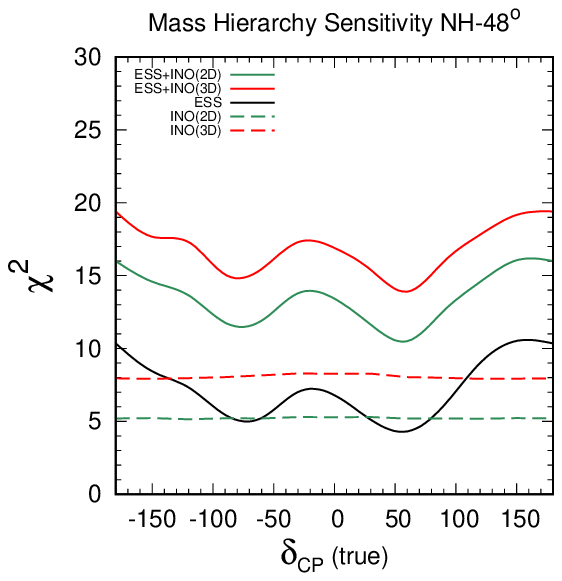}\\
         \hspace{-3cm}
         \includegraphics[width=0.55\textwidth]{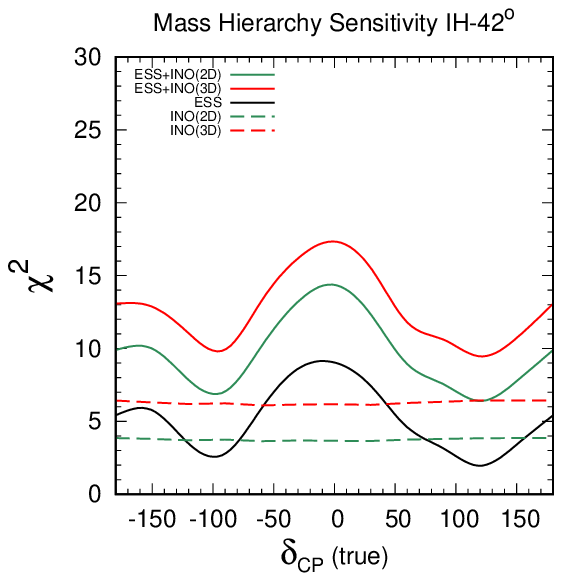}
	\hspace{-3cm}
        \includegraphics[width=0.55\textwidth]{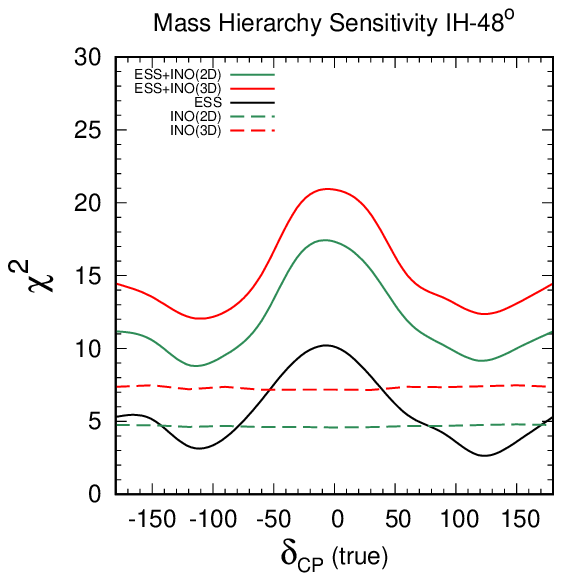}
\end{tabular}
\end{center}
\caption{Mass hierarchy sensitivity vs \dcp (true) for ICAL@INO-3D, ICAL@INO-2D and \ess for four hierarchy-octant combinations starting from top right, clockwise in the order NH-LO, NH-HO, IH-HO and IH-LO. Each figure consists of five different experimental combinations which are represented as ICAL@INO-2D (green dashed curve), ICAL@INO-3D (red dashed curve), \ess (black solid curve), \ess + ICAL@INO-2D (green solid curve) and \ess + ICAL@INO-3D (red solid curve) .}
\label{fig:MH-2D-3D}
\end{figure}

In Fig.\ref{fig:MH-2D-3D}  we compare and quantify
the hierarchy sensitivity of 
\ess + ICAL@INO for 2D and 3D analysis of the ICAL@INO experiment. 
The 3D analysis gives better hierarchy 
sensitivity over 2D because of the inclusion of 
hadron information in the analysis as described in Sec.\ref{sec:Experimental-details}.

\subsection{Octant sensitivity}
\begin{figure}[h!]
\begin{center}
        \begin{tabular}{cc}
        \hspace{-3cm}
        \includegraphics[width=0.55\textwidth]{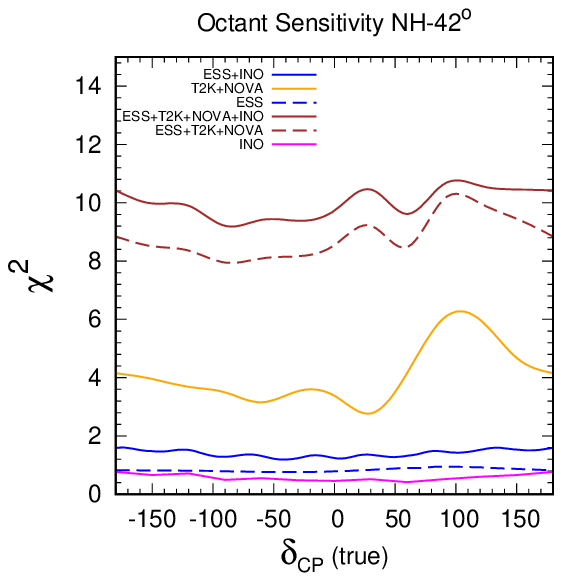}
	\hspace{-3cm}
	\includegraphics[width=0.55\textwidth]{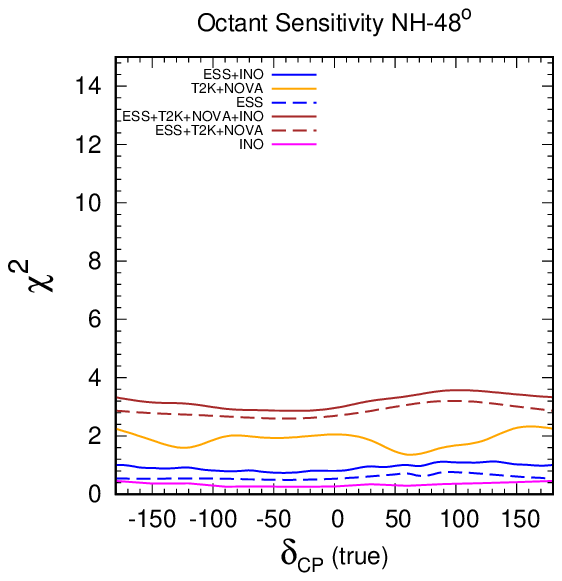}\\
         \hspace{-3cm}
         \includegraphics[width=0.55\textwidth]{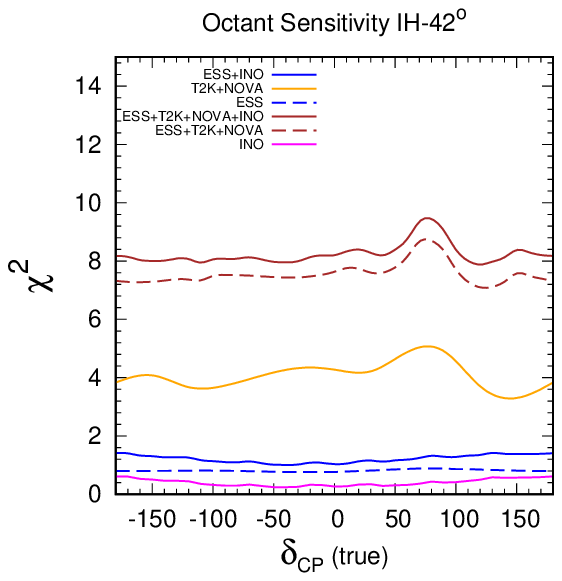}
	\hspace{-3cm}
	\includegraphics[width=0.55\textwidth]{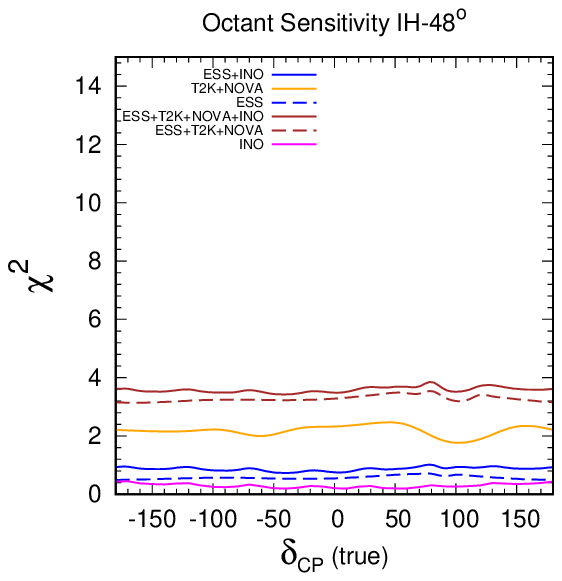}
	\end{tabular}
	\end{center}
\caption{Octant sensitivity vs \dcp (true) for ICAL@INO-3D \ess T2K \nova for four hierarchy-octant combinations starting from top right, clockwise in the order NH-LO, NH-HO, IH-HO and IH-LO. Each figure consists of six different experimental combinations which are represented as ICAL@INO (magenta solid curve), \ess (blue dashed curve), T2K + \nova (orange solid curve), \ess + ICAL@INO (blue solid curve), \ess + T2K + \nova (brown dashed curve) and \ess + T2K + \nova + ICAL@INO (brown solid curve).}
\label{fig:Octant}
\end{figure}

\begin{figure}[h!]
\begin{center}
        \vspace*{-2cm}
        \begin{tabular}{cc}
        \hspace{-3cm}
        \includegraphics[width=0.55\textwidth]{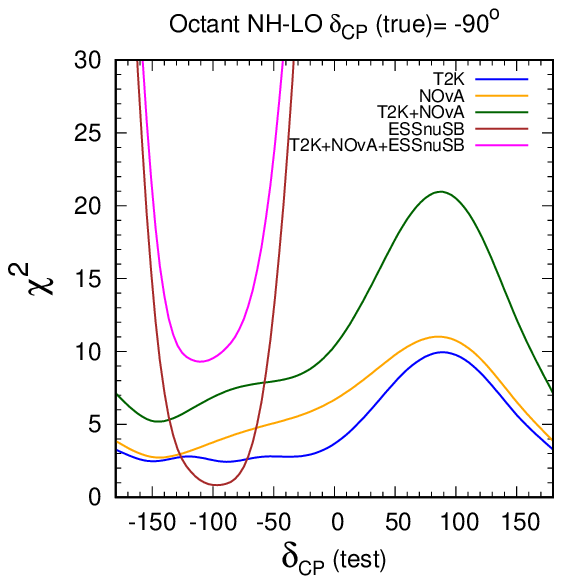}
	\hspace{-3cm}
	\includegraphics[width=0.55\textwidth]{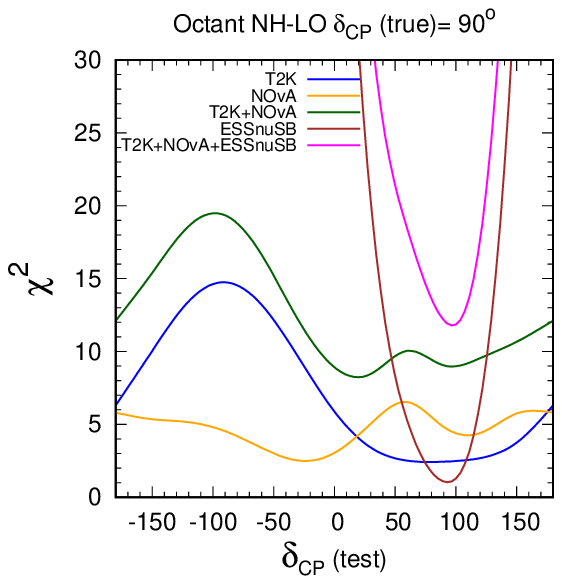}\\
	\end{tabular}
\caption{Octant sensitivity vs \dcp (test) for \ess , T2K and \nova with \dcp (true) $=\pm 90^\circ$. The true hierarchy and true octant are taken as NH and LO respectively. Left panel represents the octant sensitivities for \dcp (true) $= - 90^\circ$ and the right panel \dcp (true) $= + 90^\circ$. Each figure consists of five different experimental combinations which are represented as T2K (blue solid curve), \nova (orange solid curve), \ess (brown solid curve), T2K + \nova (green solid curve), \ess + T2K + \nova (magenta solid curve).}
\label{fig:ess-Octant-understand}
\end{center}
\end{figure}

To calculate the octant sensitivity we simulate the data for a representative
value of true $\theta_{23}$ belonging to  
LO (HO) and test it by varying $\theta_{23}$ in the 
opposite octant i.e. HO (LO) along with marginalization over 
$|\Delta m^2_{31}|$, hierarchy and \dcp (for LBL experiments). 
The plots in Fig.\ref{fig:Octant} show the octant sensitivity for the 
various experiments. 
The magenta curves denote the octant sensitivity of ICAL@INO including the hadron information.
It is seen that ICAL@INO has very poor octant sensitivity ($ \sim < 1\sigma$).
Although as discussed earlier, the matter effect can break
octant degeneracy the sensitivity is low for ICAL@INO, since it
can detect only the muon signal. This gets contribution from
both $P_{\mu\mu}$ and $P_{e\mu}$ probabilities and the octant sensitivity 
are opposite which reduces the sensitivity.

T2K, NO$\nu$A and \ess are accelerator experiments which can detect both
the appearance and disappearance channels separately.
The blue dashed line denotes the octant sensitivity of \ess which is
again $< 1\sigma$ for all the four hierarchy octant combinations.
As we have seen from the discussion on probabilities,
that \ess suffers from octant degeneracy over the whole range of
\dcp for the E1 and E3 bins. Thus, the octant $\chi^2$ gets contribution 
mainly from the bin E2. 

The solid yellow line in Fig.\ref{fig:Octant} represents the combined octant sensitivity for
T2K + NO$\nu$A. This combination has $\sim 2\sigma$ octant sensitivity for
most of the \dcp range for true NH-LO and IH-LO with peak sensitivity reaching
upto $\sim 2.4\sigma$ and $\sim 2.2\sigma$ at \dcp$\sim 90 ^\circ$ respectively. 
The combination of  neutrino and anti-neutrino run 
help in removing the degenerate  wrong octant solutions \cite{Ghosh:2015ena,Agarwalla:2013ju,Machado:2013kya,Coloma:2014kca}. 
There can also be some synergy between T2K and \nova which can enhance 
octant sensitivity. 
For instance for true NH-LO, we find a higher sensitivity in the 
upper half plane of \dcp  near \dcp$ \sim 100^\circ$. This happens 
because of synergy between T2K and NO$\nu$A.
This can be understood  from the 
Fig.\ref{fig:ess-Octant-understand} where we plot the octant sensitivity 
$\chi^2$ vs test-\dcp for T2K, \nova and \ess for NH-LO and two representative 
values of true \dcp = $\pm 90^\circ$. 
It is seen that for \dcp = $-90^\circ$ the minima of NO$\nu$A, T2K and 
the combined $\chi^2$ of \nova and T2K, represented by the blue, yellow and the 
green lines come around $-150^\circ$. 
But for \dcp =$90^\circ$ the minimum of T2K comes close to $+90^\circ$ but the 
minimum of \nova comes close to \dcp $ \sim -20^\circ$.   The 
combined minimum comes at $\sim \delta_{CP} = 20^\circ$ where the 
T2K and \nova contributions are higher than that at their 
individual minimum. This synergy gives a higher $\chi^2$ in the UHP.  

For IH-LO (green band in the probability Fig.\ref{fig:probPlots}), the neutrino probability
has no degeneracy for \dcp belonging to the upper half plane. 
The anti-neutrino probabilities for \dcp in upper half plane has 
degeneracy with NH-HO at same \dcp (purple band) 
and IH-HO with \dcp in the lower half plane. But since neutrino 
events have larger statistics, higher sensitivity is 
obtained for \dcp in the upper half plane. 
The $\chi^2$ is lower for HO in general. This is because the $\chi^2 \sim
(N_{HO(LO)} - N_{LO(HO)})/N_{HO(LO)}$
for true HO(LO). We see that the numerator is same for both cases whereas the
denominator is
larger for a true higher octant resulting in a lower sensitivity.

When T2K + \nova is combined with \ess 
(shown by the dashed brown lines in the Fig.\ref{fig:Octant}) an enhancement 
is observed with the octant sensitivity reaching $\sim 3\sigma$ at 
\dcp$\sim 90 ^\circ$ for true NH-LO
while the octant sensitivity reaches $\sim 2.9\sigma$ for true IH-LO at the same $\delta_{CP}$.
But the octant sensitivities are $\sim 1.7\sigma$ for true NH-HO and IH-HO.

It is interesting to understand the enhancement of the octant sensitivity of T2K + \nova when combined with \ess.
This synergy can be understood from Fig.\ref{fig:ess-Octant-understand}. The brown curve in this figure denotes the octant sensitivity of \ess as a function of test \dcp while the magenta 
curve denotes the combined sensitivity of T2K + \nova+ \ess. 
The left panel represents the plots with \dcp(true) = $-90^\circ$ 
while right panel represents \dcp (true) = $90^\circ$.
Analyzing the first panel i.e. NH-LO and \dcp(true) = $-90^\circ$
it is seen that the minima for both \t2k and \nova come at \dcp(test) = $-150^\circ$ hence there is no synergy between \t2k and \nova as discussed earlier.  
But, the \ess minima is at \dcp(test) = $-100^\circ$ therefore the 
overall minima is shifted towards \dcp(test) = $-120^\circ$, 
which gives rise to significant
synergy between \ess and T2K + \nova which can be seen from the 
magenta curve in Fig.\ref{fig:ess-Octant-understand}.

The variation in the octant sensitivity  $\chi^2$ with \dcp(test) is
seen to be  very rapid for \ess hence it controls the overall shape of the combined octant sensitivity curve and the position of the minima.  
As we have discussed earlier, the octant sensitivity for \ess is 
contributed by the bin with mean energy 0.35 GeV. 
As can be seen from Fig.\ref{fig:prob-ESS}, the probability for this 
bin has a sharp variation with $\delta_{CP}$. 
Thus a slight shift in the \dcp value can cause a large change in the 
probability and hence in the $\chi^2$.  
Similar feature can also be observed in the second panel for NH-LO 
and true \dcp = $90^\circ$. 

Addition of ICAL@INO with T2K + \nova + \ess represented by the 
solid brown curves,
results in slightly higher sensitivity.
In  this case for true NH-LO $3\sigma$ octant sensitivity is obtained.
For true IH-LO the octant sensitivity  reaches close to $\sim 3\sigma$ at \dcp$\sim 90 ^\circ$.
While for true NH-HO and IH-HO the total sensitivity  obtained is close to $2\sigma$.
Adding ICAL@INO, results in a constant increase in the $\chi^2$, since the ICAL@INO $\chi^2$ is almost independent of $\delta_{CP}$.

If we add priors over $\delta_{CP}$, $\Delta m^2_{31}$ and $\theta_{23}$ 
using the $3\sigma$ ranges from the global analysis of the current data
we find a 6\% increase in octant sensitivity 
due to the prior on $\delta_{CP}$. But priors over $\Delta m^2_{31}$ and $\theta_{23}$ do not play any significant role.

\section{Conclusions}\label{sec:Conclusions}

The \ess experiment is planned for discovery of \dcp with a high significance using the
second oscillation maximum. 
In this work, we show how the hierarchy sensitivity of the \ess experiment
can be enhanced by combining with the atmospheric neutrino data at the
proposed ICAL detector of the ICAL@INO collaboration as well as the
data from the ongoing T2K and \nova experiments assuming their
full projected runs.
We present our results for four true hierarchy - octant combination :
NH-LO, NH-HO, IH-LO, IH-HO taking representative values of $\theta_{23}= 42^\circ$ for LO and $48^\circ$ for HO.
We find that
\ess has $ \sim 2(3) \sigma$ hierarchy
sensitivity over the majority of \dcp values for the above hierarchy octant
combinations.
The mass hierarchy sensitivity of ICAL@INO is independent of \dcp and
adding ICAL data to \ess helps to enhance this to $3(4)\sigma$  depending
on hierarchy, octant and \dcp value.
Addition of T2K + \nova to this combination raises the hierarchy sensitivity
farther and $5\sigma$ sensitivity to mass hierarchy can be reached.
The overall conservative sensitivities (best sensitivities) for the various hierarchy octant combinations are
as follows:
\begin{itemize}
\item NH-LO : $\sim$ 4.4(5)$\sigma$
\item NH-HO : $\sim$ 5(5.5)$\sigma$
\item IH-LO : $\sim$ 4.5(5)$\sigma$
\item IH-HO : $\sim$ 4.8(5.3)$\sigma$
\end{itemize}

We have also explored to what extent the octant sensitivity of \ess can be improved by combining with
ICAL@INO and T2K and \nova simulated data. 
We find that ICAL@INO itself has very low octant sensitivity due to opposite interplay of
the survival and appearance channels nullifying the octant sensitivity.
However when T2K + \nova data is added $2(3)\sigma$ octant sensitivity can be achieved. 
Additionally, we have shown that despite the poor octant sensitivity of 
ESS$\nu$SB, it can have interesting synergy with T2K + \nova owing to 
the rapid variation of $P_{\mu e}$  with respect to $\delta_{CP}$ at the 
second oscillation maximum. Hence,  combining \ess with T2K + \nova significantly increases the combined $\chi^2$.

In conclusion, our analysis underscores the importance of exploring the synergies 
between the ongoing experiments T2K and \nova and the \ess experiment and atmospheric neutrino experiment ICAL@INO 
to give enhanced sensitivity.

\section*{Acknowledgment}
The authors acknowledge Enrique Fernandez-Martinez for his help in the simulation of ESS$\nu$SB experiment, Monojit Ghosh for discussions, S.Umashankar and Amol Dighe for useful suggestions. 
C.G. thanks the India-based Neutrino Observatory collaboration, which is jointly funded by the Department of Atomic Energy, India and Department of Science and Technology, India, for the financial support.
T.T. acknowledges support from the Ministerio de Economia y Competitividad(MINECO): Plan Estatal de Investigacion (ref. FPA2015- 65150-C3-1-P,MINECO/FEDER), Severo Ochoa Centre of Excellence and MultiDark Consolider(MINECO), and Prometeo program (Generalitat Valenciana), Spain.

\bibliographystyle{unsrt}

\bibliography{reference.bib}{}


\end{document}